\documentclass[floatfix,twocolumn,showpacs,aps,tightenlines]{revtex4}
\usepackage{graphicx}
\usepackage{color}
\usepackage{dcolumn}
\usepackage{bm}

\usepackage[mathscr]{eucal}
\usepackage{latexsym}
\usepackage{stmaryrd}

\usepackage{mathrsfs}
\usepackage{amssymb}
\usepackage{amsmath}
\usepackage{amsfonts}
\usepackage{longtable}
\usepackage{xspace}

\newcommand{\ud}{\mathrm{d}}

\newcommand{\beq}{\begin{equation}}
\newcommand{\eeq}{\end{equation}}

\newcommand{\be}{\begin{equation}}
\newcommand{\ee}{\end{equation}}
\newcommand{\bea}{\begin{eqnarray}}
\newcommand{\eea}{\end{eqnarray}}
\newcommand{\bes}{\begin{subequations}}
\newcommand{\ees}{\end{subequations}}

\usepackage{bm}
\newcommand{\twopunctures}{{\sc TwoPunctures}\xspace}
\newcommand{\carpet}{{\sc Carpet}\xspace}
\newcommand{\gena}{{\it generation 0}\xspace}
\newcommand{\genb}{{\it generation 1}\xspace}
\newcommand{\genc}{{\it generation 2}\xspace}
\newcommand{\spd}{{\it semiproper distance}\xspace}

\begin{document}

\title{Exploring the Outer Limits of Numerical Relativity}
\author{Carlos O. Lousto}
\author{Yosef Zlochower} 
\affiliation{Center for Computational Relativity and Gravitation,
School of Mathematical Sciences,
Rochester Institute of Technology, 85 Lomb Memorial Drive, Rochester,
 New York 14623}

\date{\today}

\begin{abstract} 
We perform several black-hole binary evolutions using fully nonlinear
numerical relativity techniques at separations large enough that
low-order post-Newtonian expansions are expected to be accurate.  As a
case study, we evolve an equal-mass nonspinning black-hole binary from
a quasicircular orbit at an initial coordinate separation of $D=100M$
for three different resolutions.
We find that the orbital period of this binary (in the numerical
coordinates) is $T=6422M$.  The orbital motion agrees with
post-Newtonian predictions to within $1\%$.  Interestingly, we find
that the time derivative of the coordinate separation is dominated by
a purely gauge effect leading to an apparent contraction and expansion
of the orbit at twice the orbital frequency.  Based on these results,
we improved our evolution techniques and studied a set of black hole
binaries in quasi-circular orbits starting at $D=20M$, $D=50M$, and
$D=100M$ for $\sim$ 5, 3, and 2 orbits, respectively. 
We find good agreement between the numerical results and
post-Newtonian predictions for the orbital frequency and radial decay
rate, radiated energy and angular momentum, and waveform amplitude
and phases.  The results are relevant for the future computation of
long-term waveforms to assist in the detection and analysis of
gravitational waves by the next generation of detectors as well as the
long-term simulations of black-hole binaries required to accurately
model astrophysically realistic circumbinary accretion disks.
\end{abstract}

\pacs{04.25.Dm, 04.25.Nx, 04.30.Db, 04.70.Bw} \maketitle

\section{Introduction and motivations}\label{sec:intro}

Ever since the breakthroughs in Numerical Relativity (NR) of
2005~\cite{Pretorius:2005gq, Campanelli:2005dd, Baker:2005vv},
fully-nonlinear numerical simulations of merging black-hole binary
(BHB) spacetimes have advanced at a remarkable pace. 
From a simulation perspective, BHB systems are
characterized by the mass ratio ($q=m_1/m_2$),
the individual spins of each BH ($\vec S_1, \vec S_2$),
the initial
separation $(\vec{D})$, and the momenta of each BH ($\vec P_1$, $\vec
P_2$) (or equivalently
the orbital eccentricity at some given separation).
While all BHBs
are computationally demanding, some of the interesting
corners of the above parameter space remain particularly challenging.
Initial
investigations explored the orbital dynamics of similar-mass BHBs with
low (or zero) spin, the
computation of the associated gravitational radiation, and the
characterization of final remnant products of their mergers.  These
techniques were soon expanded to study mixed systems of black holes
(BHs)
and neutron stars~\cite{Shibata:2007zm, Deaton:2013sla,
Etienne:2012te, Pannarale:2010vs}, the merger of neutron-star
binaries (forming BHs)~\cite{Baiotti:2008ra, Kiuchi:2009jt,
Anderson:2007kz}, gravitational
collapse~\cite{Baiotti:2006wm}, and multiple black hole
evolutions~\cite{Lousto:2007rj}.

Notable progress has recently been made in simulating binaries with
mass ratios as small as $q=1/100$ in~\cite{Lousto:2010ut,
Sperhake:2011ik}, highly spinning black holes (with intrinsic spins
$S/m^2\geq0.97$)~\cite{Lovelace:2011nu}, and ultra-relativistic
collisions of black holes (reaching a Lorentz factor of
$\gamma=2.9$)~\cite{Sperhake:2008ga}.

The close limit of black holes in numerical relativity
has been studied in detail by the 'Lazarus' approach
~\cite{Baker:2000zh,Baker:2001nu,Baker:2001sf,Baker:2002qf,Baker:2003ds,Campanelli:2004zw,Campanelli:2005ia}.
In this paper we perform a first study of BHBs at
relatively ``large'' (from the relativistic point of view) initial separations.
While we expect this problem to be well described by the
post-Newtonian approximations to general relativity, it is important to
assess how well the current numerical relativity techniques model
this case, and to what extent they have to be adapted or corrected
to reach a given accuracy.

One of the main applications of the computation of accurate gravitational
waveforms from BHBs is to assist gravitational wave
observatories in detecting and analyzing gravitational radiation from
astrophysical systems; merging BHBs being a primary
target~\cite{Aylott:2009ya,Ajith:2012tt}.  With the enhanced
low-frequency sensitivity of Advanced LIGO compared to initial LIGO,
 much longer waveform templates are required as binaries will stay in
the low-frequency part of the advanced LIGO sensitivity band for a much longer
period.  In addition, the more stringent accuracy criteria needed to
match full numerical waveforms to post-Newtonian (or other analytical
waveforms) \cite{Lindblom:2008cm, Lindblom:2010mh, Damour:2010zb,
Boyle:2011dy, MacDonald:2011ne, Ohme:2011zm} require up to a factor
ten more waveform cycles than is the current norm.
To explore the future possibility of generating such 
waveforms (and even longer) by purely full numerical techniques, we consider the case 
of a BHB at an initial separation of $D=100M$.

An independent motivation to study large-separation binaries comes
from magnetohydrodynamics (MHD) studies of circumbinary disks around
merging BHBs,
which require many hundreds of orbital cycles before the accreting
matter settles into a quasi-stationary state~\cite{Noble:2012xz}.
BHB evolution can either provide a background spacetime to 
evolve the MHD system, or the MHD and NR fields can 
be integrated in a self-consistent evolution.

In Sec.~\ref{sec:numsim}, we describe our numerical techniques
to evolve black hole binaries using the {\it moving puncture} approach
\cite{Campanelli:2005dd}. In Sec.~\ref{sec:firstgen}, we describe
the results of evolving a BHB starting at a separation of $D=100M$
using the standard techniques developed for much closer binaries
($D\sim10M$). We study the convergence of the orbital tracks of
the holes for three resolutions (refined by successive factors of
1.2).
We show results of new BHB simulations of $D=100M$, $D=50M$, and 
$D=20M$ BHBs using improved evolution techniques in 
Sec.~\ref{sec:secondgen}. We also include in our analysis
the proper separation between the two  BHs,
its decay, and 
the gravitational waveforms extraction at (at least) one wavelength from the
sources. We compare all results with 3.5 post-Newtonian predictions.
We conclude
in Sec.~\ref{sec:discussion} that current full numerical techniques
are able to accurately simulate the large separation regime
but that the much longer time scales involved in the problem
means that increases in simulations speeds by one to two orders
of magnitude will be needed to complete the evolutions to merger.

\section{Numerical Simulations}\label{sec:numsim}

We perform two different types of simulations here, a family of
lower-accuracy, but high-speed, simulations of a BHB with initial
separation of $D=100M$ at different resolutions, and a family of
higher-accuracy simulations at a fixed resolution but with different
initial separations. Note, here and below we use $M$ to denote the
unit of distance, mass, energy, etc., while we use
$m$ to denote the mass of an object. To avoid confusion between the
mass of a BHB and the mass of its constituent BHs, we denote the
masses of each BH in a binary with a subscript.

We use the TwoPunctures thorn~\cite{Ansorg:2004ds} to generate initial
puncture data~\cite{Brandt97b} for the BHB simulations described
below. These data are characterized by mass parameters $m_p$ (which
are not the horizon masses), as well as the momentum and
spin,  of each BH.  We evolve these BHB data sets using the {\sc
LazEv}~\cite{Zlochower:2005bj} implementation of the moving puncture
approach~\cite{Campanelli:2005dd,Baker:2005vv} with the conformal
function $W=\sqrt{\chi}=\exp(-2\phi)$ suggested by
Ref.~\cite{Marronetti:2007wz}.  For the runs presented here, we use
centered, eighth-order finite differencing in
space~\cite{Lousto:2007rj} and a fourth-order Runge Kutta time
integrator. (Note that we do not upwind the advection terms.) Our code
uses the {\sc Cactus}/{\sc EinsteinToolkit}~\cite{cactus_web,
einsteintoolkit, Loffler:2011ay} infrastructure.  We use the {\sc
Carpet}~\cite{Schnetter-etal-03b} mesh refinement driver to provide a
``moving boxes'' style of mesh refinement. In this approach refined
grids of fixed size are arranged about the coordinate centers of both
holes.  The {\sc Carpet} code then moves these fine grids about the
computational domain by following the trajectories of the two BHs.

To reduce the computational costs, we performed an initial set of
simulations (denoted by \gena below) that use several low-accuracy
approximations.  Among them are the techniques introduced in
Ref.~\cite{Brugmann:2008zz} where the number of buffer zones at AMR
boundaries is reduced by lowering the order of finite differencing by
successive orders near the AMR boundaries (here we use 6 buffer zones), the use of simple
interpolations of spectral initial data rather than using the complete
spectral expansion~\cite{Ansorg:2004ds}, and  the copying the initial
data to the two past time levels for use in prolongation at the
initial timestep. All of these approximation proved to be useful for
reducing the cost of numerical simulations, but each one also has the
side-effect of introducing a (hopefully) small ${\cal O}(h)$ error. 

For our higher-accuracy simulations (denoted by \genb below),
 we use a full complement of 16
buffer zones at refinement boundaries, use the full \twopunctures
spectral expansion to generate initial data, and use the \carpet {\tt
init\_3\_timelevels} option to populate the $t<0$ timelevels required
for prolongation. We also used a strict 2:1 refinement in time, which
is required for the {\tt
init\_3\_timelevels} option. In addition, we placed the outer
boundaries at $\sim 3000 M$, far enough to obtain waveforms.

To compare the run speeds with and without the lower-order
approximations, we performed two small runs using the grid structure
and CFL factors of the \genb simulations, one with
the full complement of 16 buffer zones, and the other with only 
6 buffer zones. The
latter was faster by 28\% (136 M /day compared to 106 M /day)
The tests were performed  on 20 nodes consisting of
dual 6-core 3.47 GHz Nehalem CPUs. Increasing the CFL factor increases
the run speed correspondingly (hence these low-accuracy techniques are
quite useful in situations where high-accuracy is not required).

We use a modified 1+log lapse and a modified $\Gamma$-driver
shift
condition~\cite{Alcubierre02a,Campanelli:2005dd,vanMeter:2006vi}, and
an initial lapse $\alpha(t=0) = 2/(1+\psi_{BL}^{4})$, where
$\psi_{BL}$ is the Brill-Lindquist conformal factor and is given by $$
\psi_{BL} = 1 + \sum_{i=1}^n m_{i}^p / (2 |\vec r- \vec r_i|), $$
where $\vec r_i$ is the coordinate location of puncture $i$.  The
lapse and shift are evolved with \begin{subequations} \label{eq:gauge}
\begin{eqnarray} (\partial_t - \beta^i \partial_i) \alpha &=& - 2
\alpha K,\\ \partial_t \beta^a &=& (3/4) \tilde \Gamma^a - \sigma
\beta^a \,, \label{eq:Bdot} \end{eqnarray} \end{subequations}
 where we
use $\sigma=2$ for the \gena simulations presented below, while
we use a spatially varying $\sigma = (\sigma_0 - \sigma_\infty)
\exp(-r^4/w^4)+ \sigma_\infty$ (where $\sigma_0=2.0$,
$\sigma_\infty=0.25$, $w=80.0$) for the \genb simulations.
The reason for the differences is that, since we use strict 2:1
refinement in time for \genb, we need a small $\sigma$ at large distances in
order for the system to remain stable~\cite{Schnetter:2010cz}.

We use {\sc AHFinderDirect}~\cite{Thornburg2003:AH-finding} to locate
apparent horizons. Here the spins of the two BHs are negligible
and the horizon mass is equal to the irreducible mass (i.e.\ $m_{\rm irr} =
\sqrt{A/(16 \pi)}$, where $A$ is the surface area of
the horizon).

\section{Zeroth Generation Runs}\label{sec:firstgen}

Here we use our standard full numerical techniques
(developed to evolve BHB at initial separations $D\sim10M$) to
evolve a BHB at a separation of $D=100M$ and
compare with the post-Newtonian predictions for the orbital
trajectory.

\subsection{Quasicircular Post-Newtonian Analysis}

BHB quasicircular orbits at separations $r\sim100M$ (note that we use
$r$ to denote the post-Newtonian (PN) expression for the orbital separation)
 should be described
 accurately by
the 2PN~\cite{Kidder:1995zr} expressions
for the orbital frequency
\begin{equation}\label{eq:Omega}
\Omega^2=\frac{m}{r^3}\left[1-(3-\eta)\,\frac{m}{r}+
\left(6+\frac{41}{4}\eta+\eta^2\right)\,\frac{m^2}{r^2}\right]+\cdots
\end{equation}
and radial decay of the orbit (due to energy loss)
\begin{equation}\label{eq:rdot}
\dot{r}=-\frac{64M^3}{5r^3}\eta
\left[1-\frac{(1751+588\eta)}{336}\,\frac{m}{r}+4\pi\,\left(\frac{m}{r}\right)^{3/2}\right]+\cdots,
\end{equation}
where $m=m_1+m_2$ is the total mass and $\eta=(m_1 m_2)/m^2$.
The orbital period, $T$, at $r=100M$ for an equal-mass BHB
 is given by
\begin{equation}\label{eq:T}
T=\frac{2\pi}{\Omega}\approx6369M,
\end{equation}
where $m\Omega\approx0.001$, and $m=1M$.

We thus estimate that the radial decay after one orbit is
\begin{equation}\label{eq:Deltar}
\Delta\,r=\dot{r}\,T\approx-0.02M.
\end{equation}

Note that in the above expressions
$r$ is the harmonic radial coordinate, which
is related to the ADM-TT coordinate $R$ 
(initially related to the numerical coordinates) by
\footnote{Provided by H.Nakano based on \cite{Damour:2000ni}}
$r=R+m^2(2+29\eta)/(8R)+\cdots$.
We also neglect the effects of the small eccentricities 
that are present in the full numerical runs.

\subsection{$D=100M$ Runs versus resolution}

In order to generate initial data, we need the puncture (coordinate)
separation and momenta. We obtain approximations for the separation
and momenta by using the associated 3.5PN parameters.
Our procedure is as follows: We start with 3.0 PN quasi-circular
parameters for an orbital separation of $D\sim100M$. We then perform
a 3.5PN evolution~\cite{Buonanno:2005xu, Damour:2007nc,
Steinhoff:2007mb} of the orbital motion. The resulting orbit will be
(moderately) eccentric. We then use the procedure
in~\cite{Pfeiffer:2007yz} to obtain successively better approximations
of the initial momenta for this 3.5PN evolution 
until the eccentricity is reduced to the roundoff 
limit. After renormalizing
to an ADM mass of $1M$, we use the resulting PN parameters to solve
for the initial data. The parameters are given in Table~\ref{tab:ID}.
\begin{table}[t]
\caption{Initial data parameters for the full numerical simulations. The punctures are located at $\pm(x,0,0)$, with momenta $\pm(p_x,p_y,0)$, spin
$(0,0,0)$, and puncture mass $m_p$. $m_H$ is the measured horizon masses,
while $T_i$ is the period of the first orbit for the three resolution
runs (0 for coarsest,1 for medium, 2 for finest).
$m_{\rm PN}$ and $T_{\rm PN}$ are the corresponding post-Newtonian values}
\label{tab:ID}
\begin{ruledtabular}
\begin{tabular}{lll}
$x=50$ & $T_{\rm PN}=6365$\\
$p_x=-7.97939\times10^{-7}$ & $T_0=6406$ \\
$p_y=2.55526\times10^{-2}$ & $T_1=6420$ \\
$m_p=0.49920645$ & $T_2=6422$ \\
$m_H= 0.500615$ & $m_{\rm PN}=0.500616$\\
\end{tabular}
\end{ruledtabular}
\end{table}

We evolved the binary at three different resolutions 
characterized by a coarse-level resolution
of $h_0=4M$, $h_1=4M/1.2$, and $h_2=M/1.44$, respectively.
For these runs, we choose a Courant-Friedrichs-Lewy (CFL) factor
(i.e.\ the ratio $dt/h$ between the
timestep and gridsize) that  varies with refinement level, as described in
Table~\ref{tab:levels}. 
In all cases, the boundaries were located at $400M$ and 9 levels of
(2:1) refinement were used.  This means that the boundary
is causally connected to the binary for most of the simulation.
\begin{table}[t]
\caption{The grid structure for the three numerical simulations
presented here. The coarsest run had $h=h_0=4M$, while the
two higher resolution runs had $h=h_1=4 M/1.2$ and
 $h=h_2=4M/1.44$, respectively.} \label{tab:levels}
\begin{ruledtabular}
\begin{tabular}{l|lll}
& radius & resolution & CFL\\
\hline
0 & 400 & $h$ & 0.019 \\
1 & 220 & $h/2$ & 0.038 \\
2 & 110 & $h/4$ & 0.076 \\
3 & 55  & $h/8$ & 0.152 \\
4 & 25  & $h/16$ & 0.152 \\
5 & 10  & $h/32$ & 0.304 \\
6 & 5   & $h/64$ & 0.304 \\
7 & 2   & $h/128$ & 0.304 \\
8 & 0.65 & $h/256$ & 0.304 \\
\end{tabular}
\end{ruledtabular}
\end{table}

\begin{figure}
  \includegraphics[width=\columnwidth]{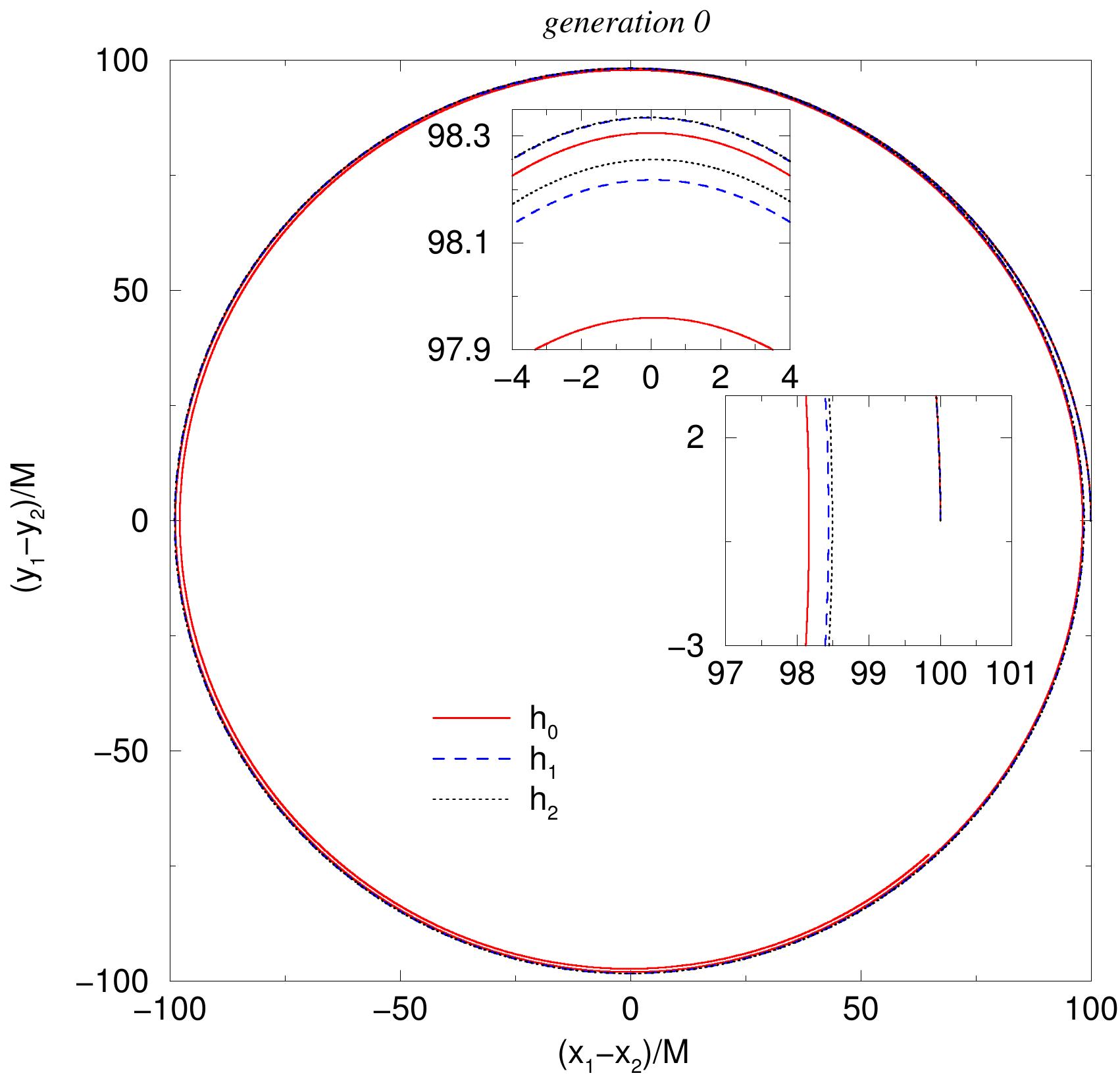}
  \caption{The orbital trajectory for
the \gena $D=100M$ simulations at different resolutions.}
\label{fig:fermium_track}
\end{figure}

\begin{figure}
  \includegraphics[width=\columnwidth]{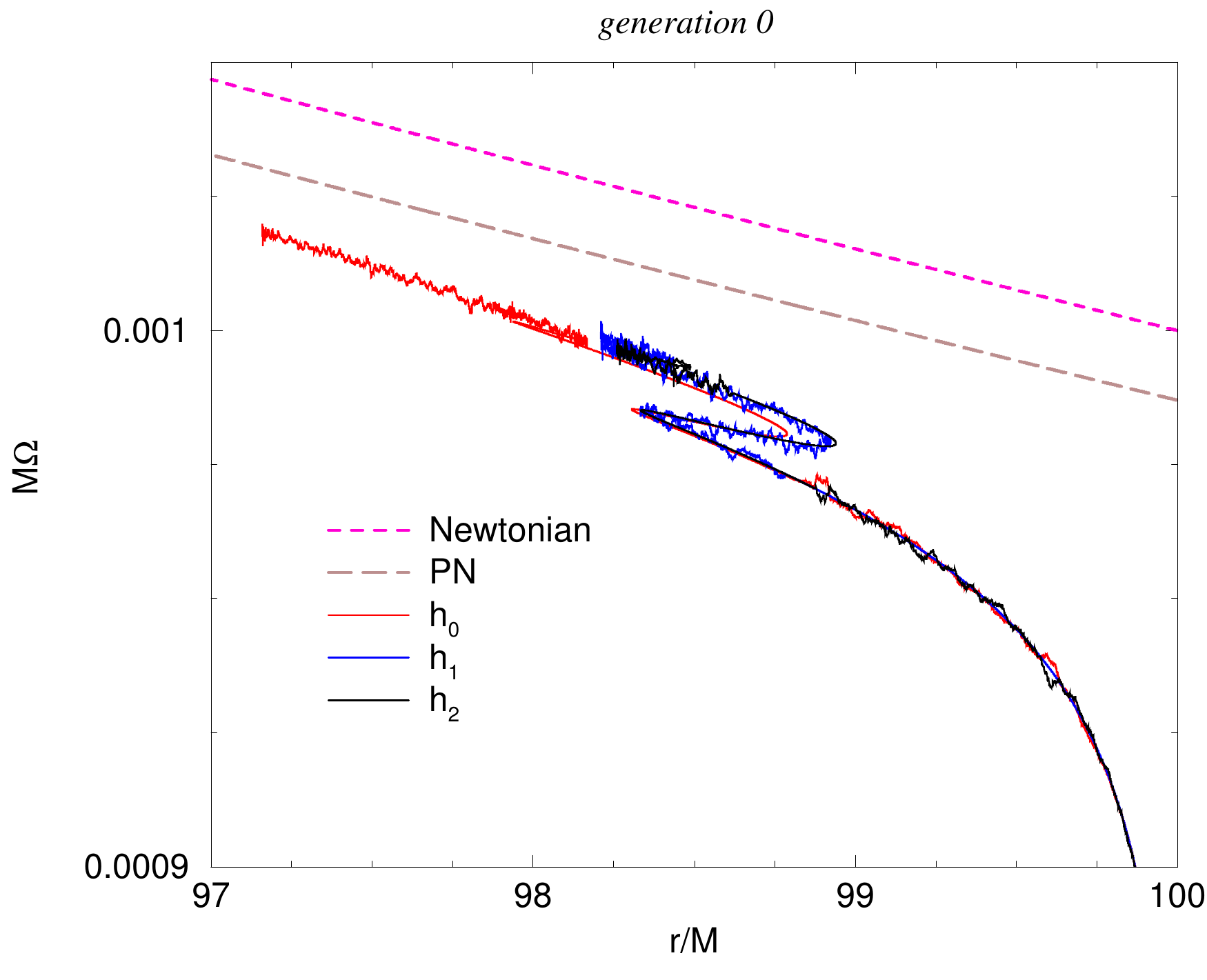}
\caption{The orbital frequency $d\phi_{\rm orbit}/dt$ versus orbital
separation $r$, as well as the Newtonian and 3.5 PN predictions for
the \gena $D=100M$ simulations. The
mapping between $r$ and $\Omega$ is not unique because $r$ is not a
monotonic function of $t$.}
 \label{fig:fermium_freq}
\end{figure}

As seen in Fig.~\ref{fig:fermium_track},
the trajectory of the full numerical
simulation is  almost a
closed circle with radius $r\sim100M$.
A more quantitative comparison is shown in Fig~\ref{fig:fermium_freq},
where we plot the
orbital frequency as a function of orbital separation and compare to
the 3.5PN predictions. As seen in Fig~\ref{fig:fermium_freq}, the
orbital frequency starts out much lower than predicted (which is
purely a gauge effect) and then approaches the PN value from below. 


\begin{figure}
  \includegraphics[width=\columnwidth]{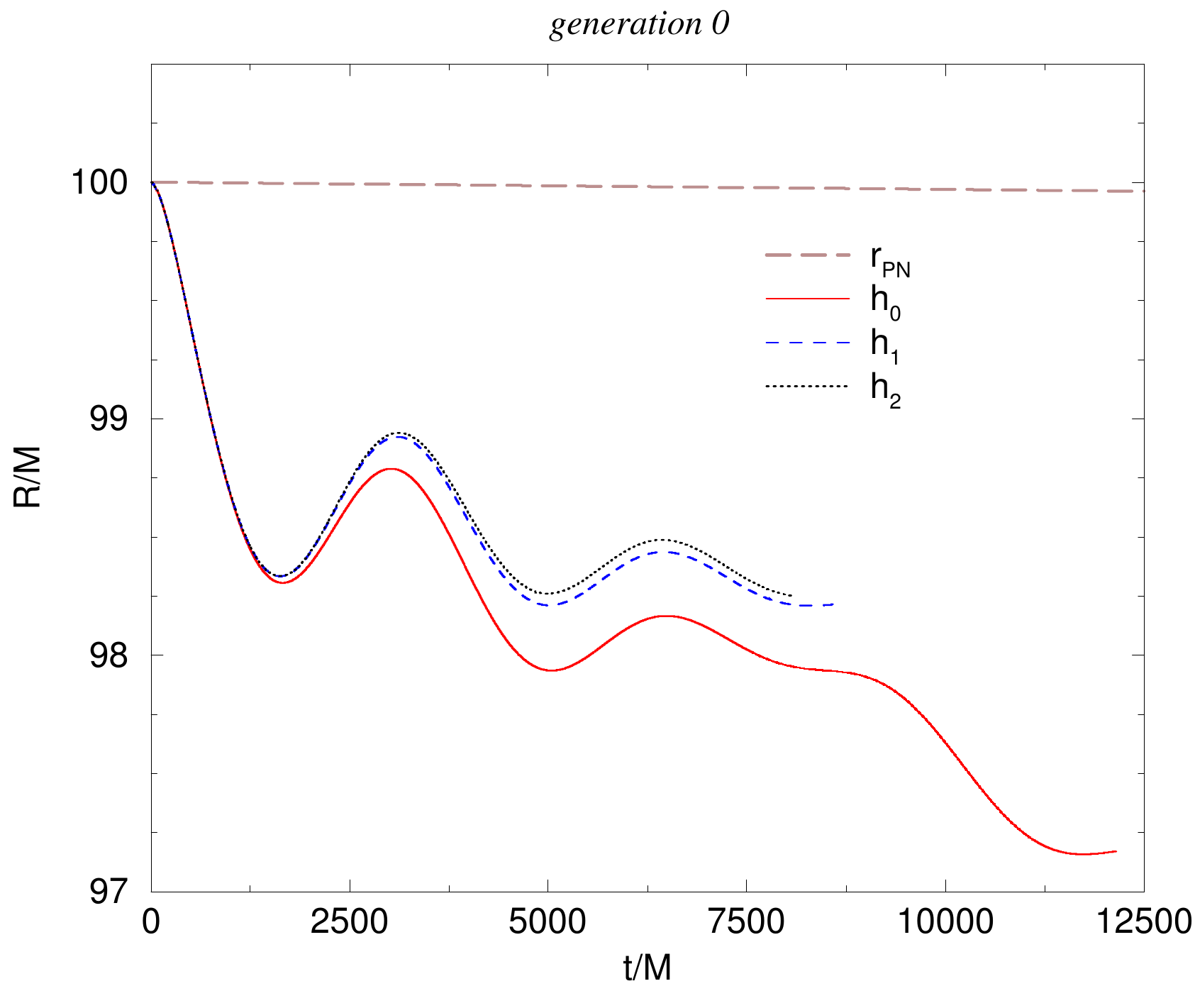}
  \caption{The orbital separation versus time at three resolutions, as
well as the 3.5PN prediction, for the \gena $D=100M$ simulations.}
\label{fig:coord_r}
\end{figure}

\begin{figure}
  \includegraphics[width=\columnwidth]{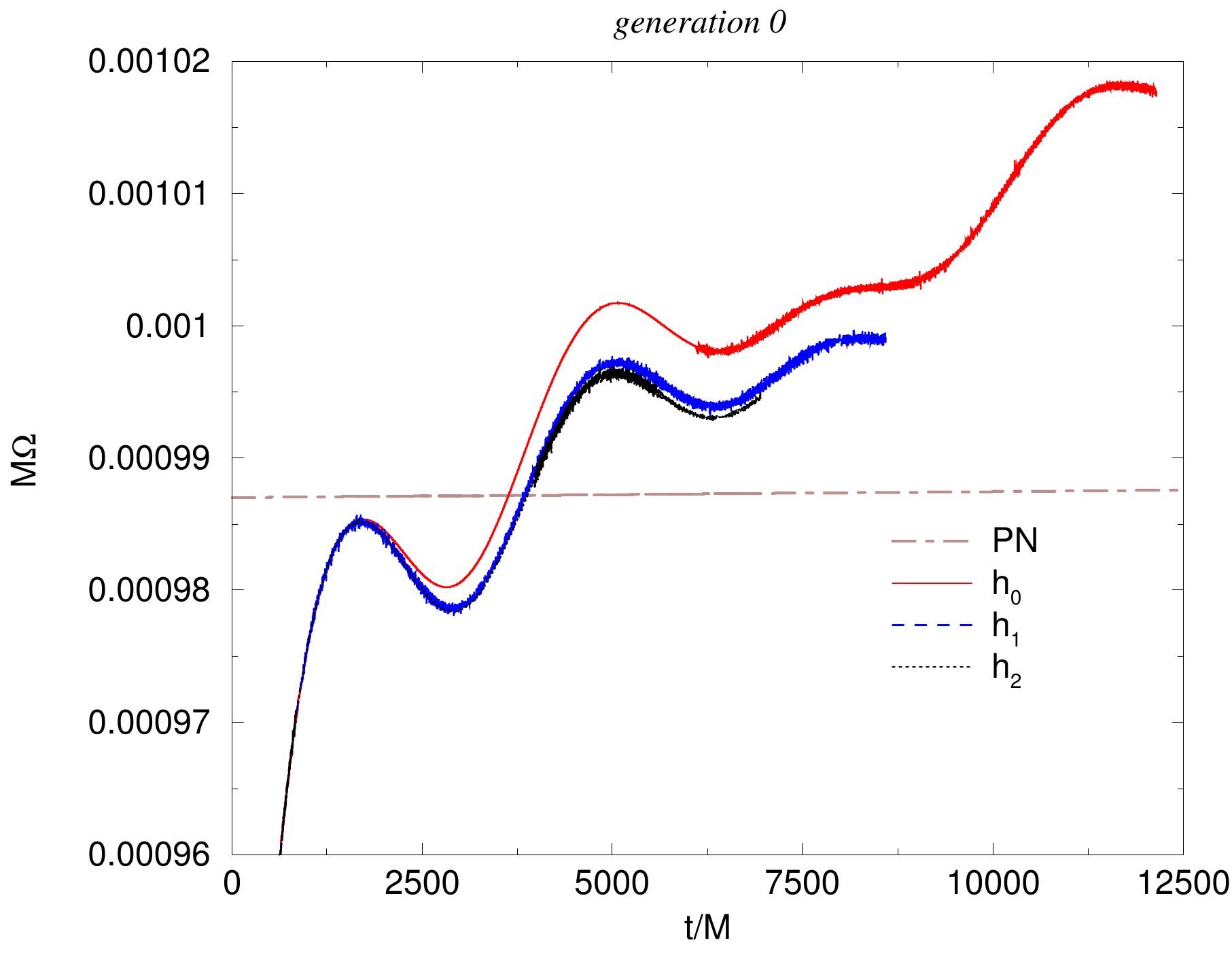}
  \caption{The orbital frequency versus time at three resolutions, as
well as the 3.5PN prediction, for the \gena $D=100M$ simulations..}
\label{fig:coord_omega}
\end{figure}

A closer look at the time dependence of the coordinate separation
(and its time derivative) shows the limited accuracy of the runs.
A pure 3.5PN evolution indicates that this binary will take
approximately $t\sim 8.2\times10^6M$ to inspiral to an orbital
separation of $r=5M$. 
During the first orbit, the 3.5PN binary inspirals only
$0.02M$ from the initial separation of $r=100M$. As shown below, the
full numerical binaries inspiral by two orders of magnitude more.  While
the 3.5PN simulation indicates that this binary is not eccentric,
there are significant oscillations in the orbital radius (usually
associated with eccentricity). Interestingly, these oscillations occur
at twice the orbital frequency (see
Fig.~\ref{fig:coord_r}), while for eccentricity effects 
we expect oscillations at the orbital frequency.
In Figs.~\ref{fig:coord_r}~and~\ref{fig:coord_omega}, we plot the
orbital separation and frequency versus time and compare to the
PN prediction.
Again, a significant difference between the PN and
numerical results is apparent. We will show in the next section
that this behavior is mostly a coordinate artifact and proper distance 
measurements greatly ameliorate this issue. The coordinate artifact
may be produced by the zero speed gauge mode \cite{vanMeter:2006vi}
contained in our choice of the shift condition (\ref{eq:Bdot}).


\begin{figure}
  \includegraphics[width=\columnwidth]{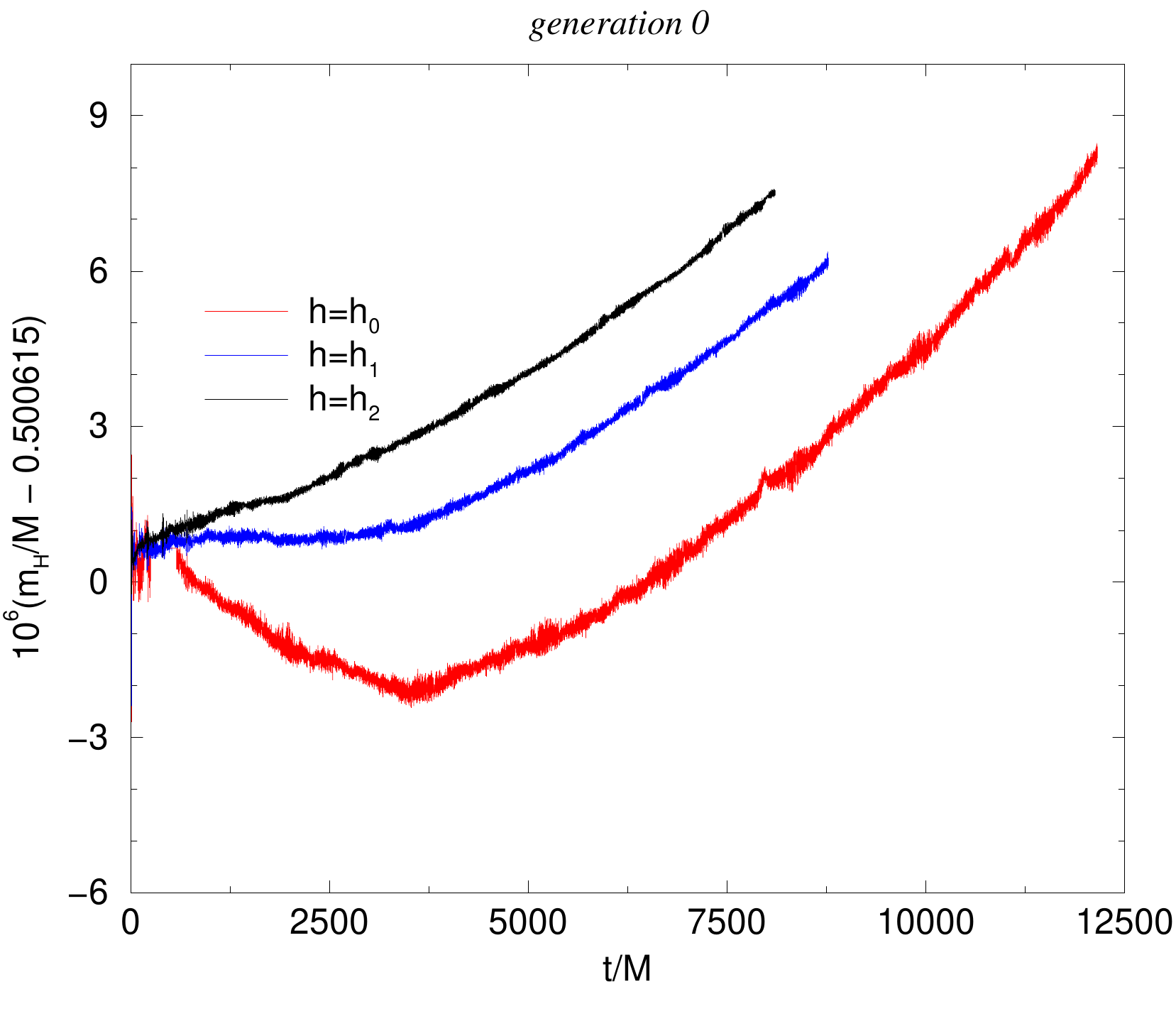}
  \caption{The horizon mass of the individual BHs 
versus time of the \gena simulations for the three resolutions.
Both the secular growth with time for all resolution, and the 
growth in mass as the gridspacing is reduced,
indicate a loss of
accuracy at later times for these \gena simulations.}
\label{fig:fermium_mass}
\end{figure}

In a previous study~\cite{Nakano:2011pb}, we found that unphysical
increases or decreases in the horizon mass can have a significant
impact on the orbital trajectory. In Fig.~\ref{fig:fermium_mass} we
plot the horizon mass as a function of resolution versus time. We
observe an increase in mass, which may be the cause of the artificial
inspiral. The mass increase is small, but this secular 
growth (which gets worse as the resolution is increased)
can affect the accuracy of longer term runs.
We will see how to better control this artifact in the next section.

\begin{figure}
  \includegraphics[width=0.49\columnwidth]{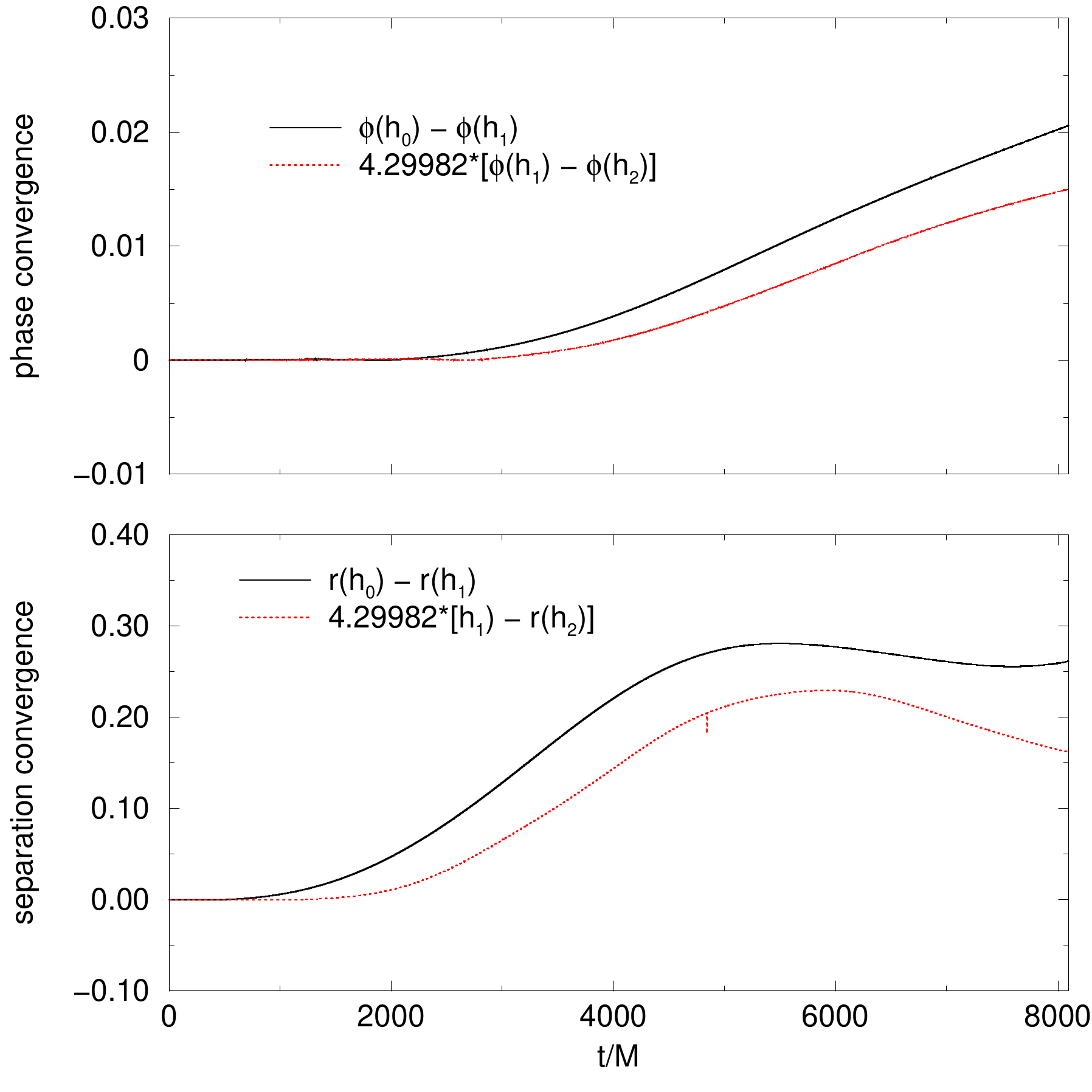}
  \includegraphics[width=0.49\columnwidth]{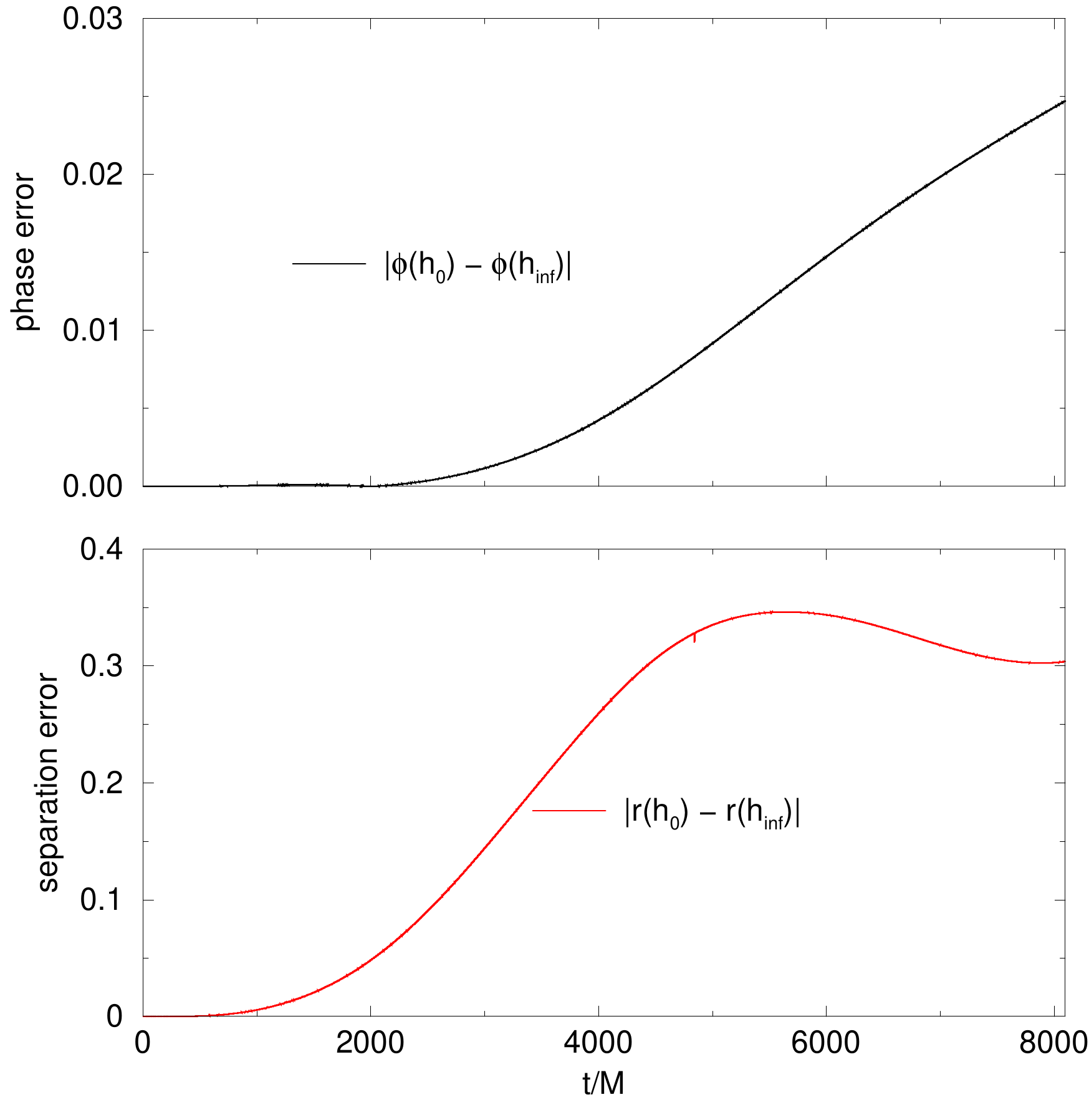}
\caption{ (Left)  Convergence plots for the orbital phase (top-left)  
and separation (bottom-left)
showing stronger than eighth-order convergence.
If a quantity $F(h)$ is eighth-order convergent, then
$F(h_0)  - F(h_1) = 4.29982 [ F(h_1) - F(h_2)]$. Here we see {\it
hyperconvergence}, which almost certainly means that the errors is
oscillating with $h$ rather than behaving monotonically.
The top-right plot shows the phase error of the coarsest
resolution run (assuming eighth-order convergence and comparing the 
Richardson extrapolated value), while the
bottom-right plot shows the orbital separation error.
}
\label{fig:ferm_conv}
\end{figure}

Finally, as seen in Fig.~\ref{fig:ferm_conv},
the convergence order, as measured by comparing
the differences in the orbital separations  and phase 
between the $h_0$ and
$h_1$ runs and the $h_1$ and $h_2$ runs, appears to be higher than
eighth order. 
In the past, when we encountered this apparent super
convergence, we found in the end that the error was an oscillatory
function of the grid resolution (see~\cite{Zlochower:2012fk}). 
However, if we attempt to extrapolate, assuming eighth-order
convergence, we find that the error in the orbital separation
increases to about $0.34 M$ and then appears to level off. The phase
error increases to about $0.025$ rad.

\section{First Generation Runs}\label{sec:secondgen}

In our study of waveform accuracy for closer BHBs~\cite{Zlochower:2012fk}, 
we found that the low-accuracy approximations used in our \gena
simulations may have a larger impact than expected.  As seen in
Fig.~\ref{fig:fermium_mass}, mass gain is an issue for \gena.
By {\it correctly} initializing the grid and 
 by lowering the CFL factor to $0.25$, we mitigate this mass loss
issue. Importantly, we also place the outer boundaries a factor of
$\sim 10$ farther away. Note that it is the CFL factor in the inner
regions (near each BH) that is responsible for the conservation (or
lack thereof) of the BH mass. So even though the \genb runs have a
larger CFL factor in the outer zones then the \gena runs, they are still more
accurate in terms of mass preservation.

As an additional analysis tool, we use the proper distance between the
two horizons as measured along the coordinate line joining the two
punctures~\cite{Alcubierre:2004hr}, which we call the \spd,
or SPD, below (note that the minimal
geodesic does not follow this line).

For the \genb  runs, we performed simulations with 
a grid structure similar to the low resolution $h_0$ \gena run,
but with several additional coarser levels, and
studied BHBs with initial orbital separations of $D=20M$, $D=50M$,
 and $D=100M$. We placed the outer
boundaries at $3200M$ and used 14 levels of refinement with a coarsest
resolution of $h=32M$ (extending the grid to $6400M$ stably can be achieved by 
halving the gauge baseline $\sigma_\infty$ or extending the grid
without refining it by a factor 2 or halving the CFL condition).
We used strict 2:1 refinement in time and a
full complement of 16 buffer zones at the boundary of each refinement
level.

\begin{figure}
  \includegraphics[width=\columnwidth]{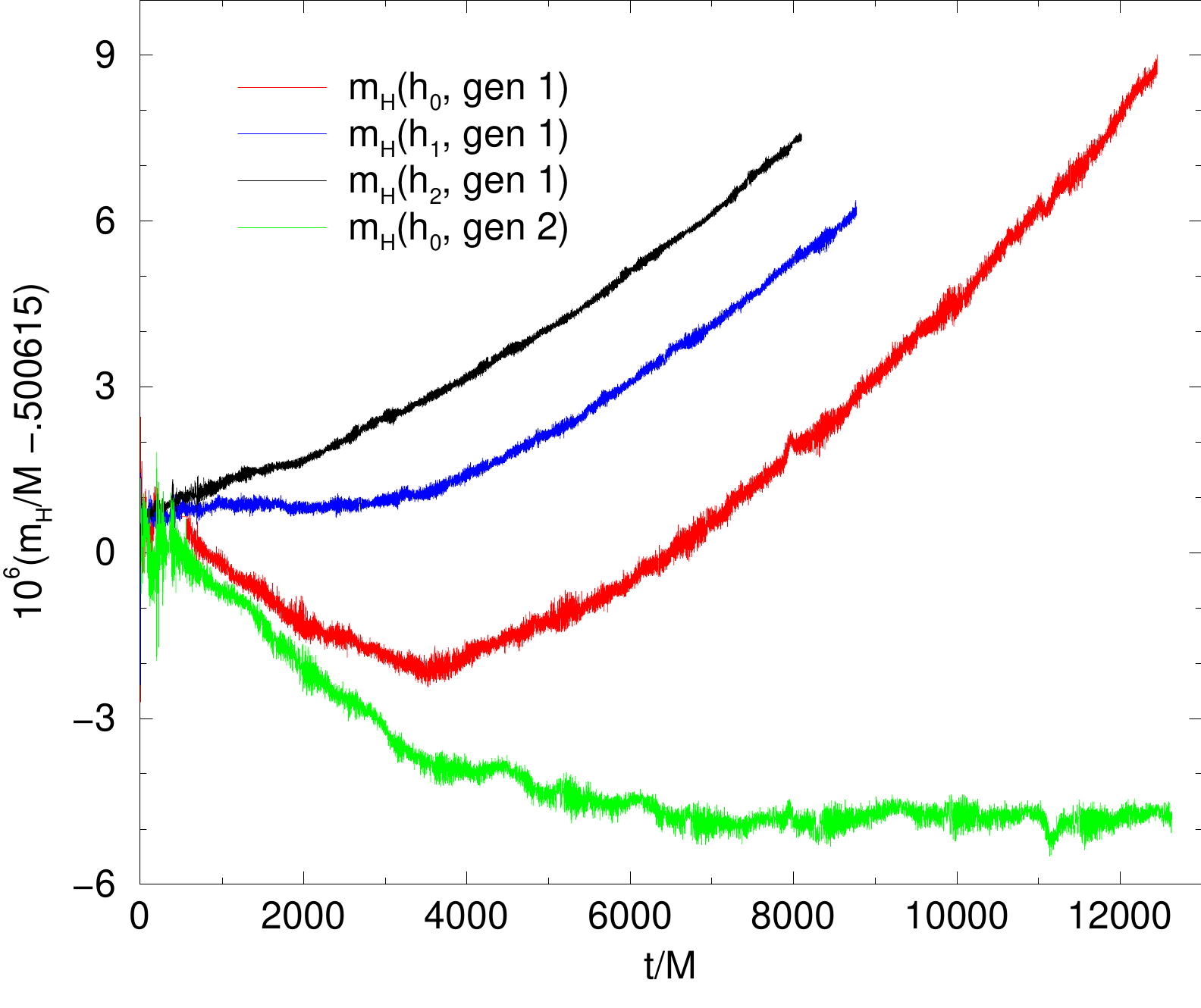}
  \caption{Comparing the $D=100M$ mass
conservation for the \gena and \genb runs. Note that the \genb simulation shows no late time growth. 
The conservation of the mass at levels of below 1 part in $10^5$
is crucial for long term evolutions.}
\label{fig:horizon_mass}
\end{figure}

The conservation of the BH masses during the evolution
is an important indicator of the accuracy of the full numerical runs,
and is particularly important for long term evolutions.
We show the horizon mass for the $D=100M$ \genb run and compare with the
\gena runs in Fig.~\ref{fig:horizon_mass}.
We observe that while \gena runs
show an strong growth of the masses with time and with
increasing resolution, the profile of the \genb  runs is rather
flat, and we expect this to be still the case for higher resolutions
as observed in Ref.~\cite{Zlochower:2012fk}, Fig.~2.

\subsection{Varying the BHB initial separation}

\begin{figure}
  \includegraphics[width=\columnwidth]{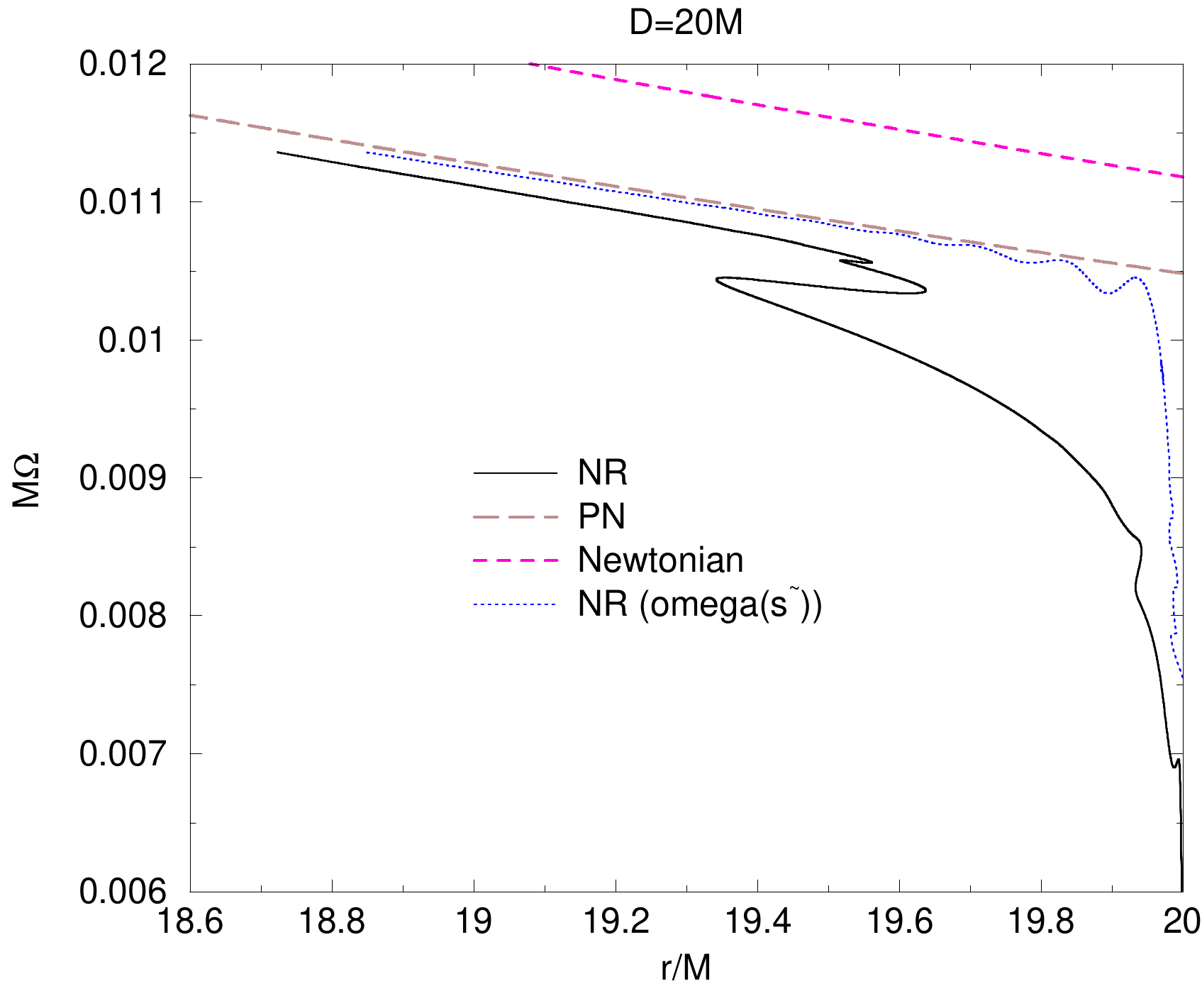}
  \includegraphics[width=\columnwidth]{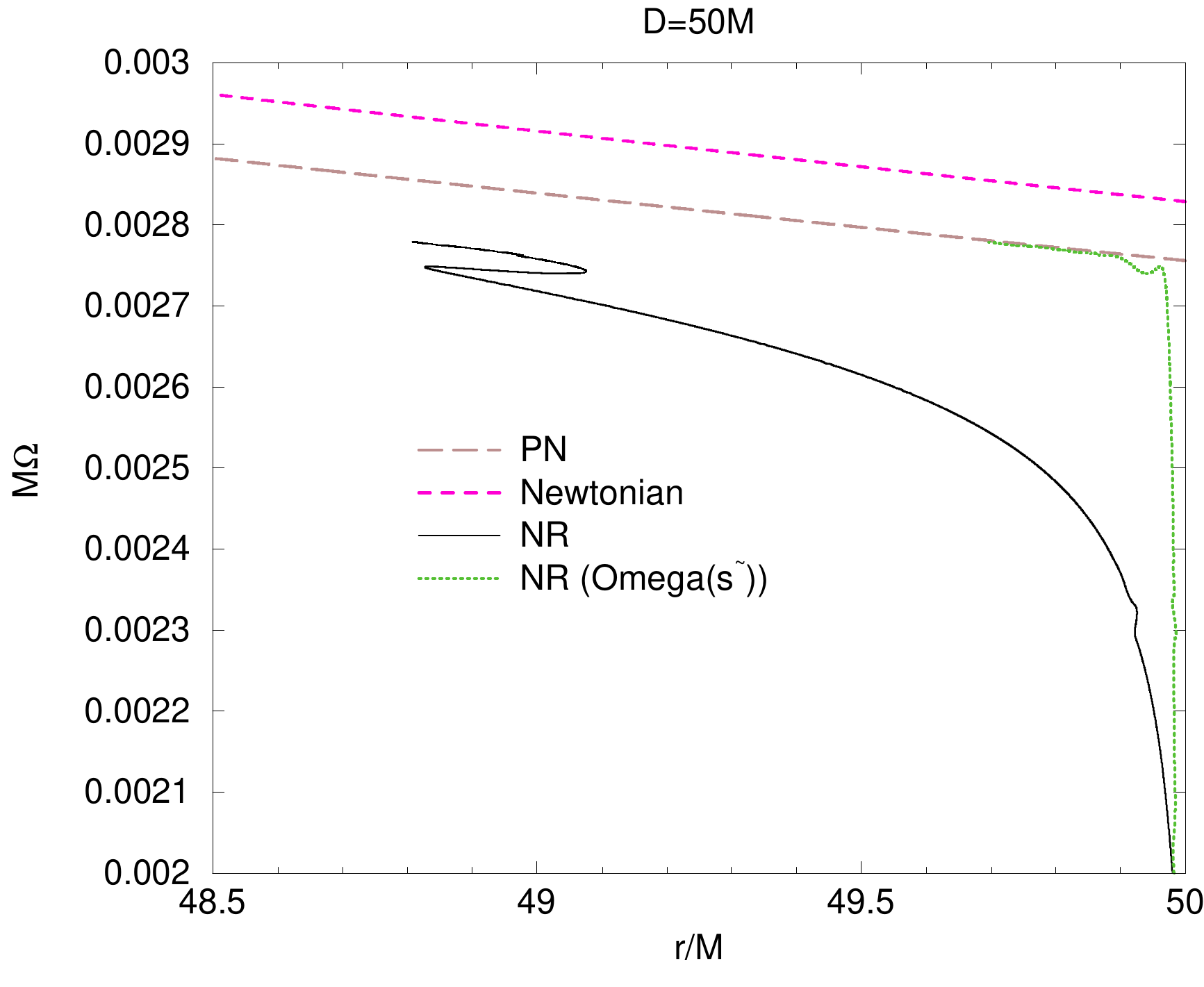}
  \includegraphics[width=\columnwidth]{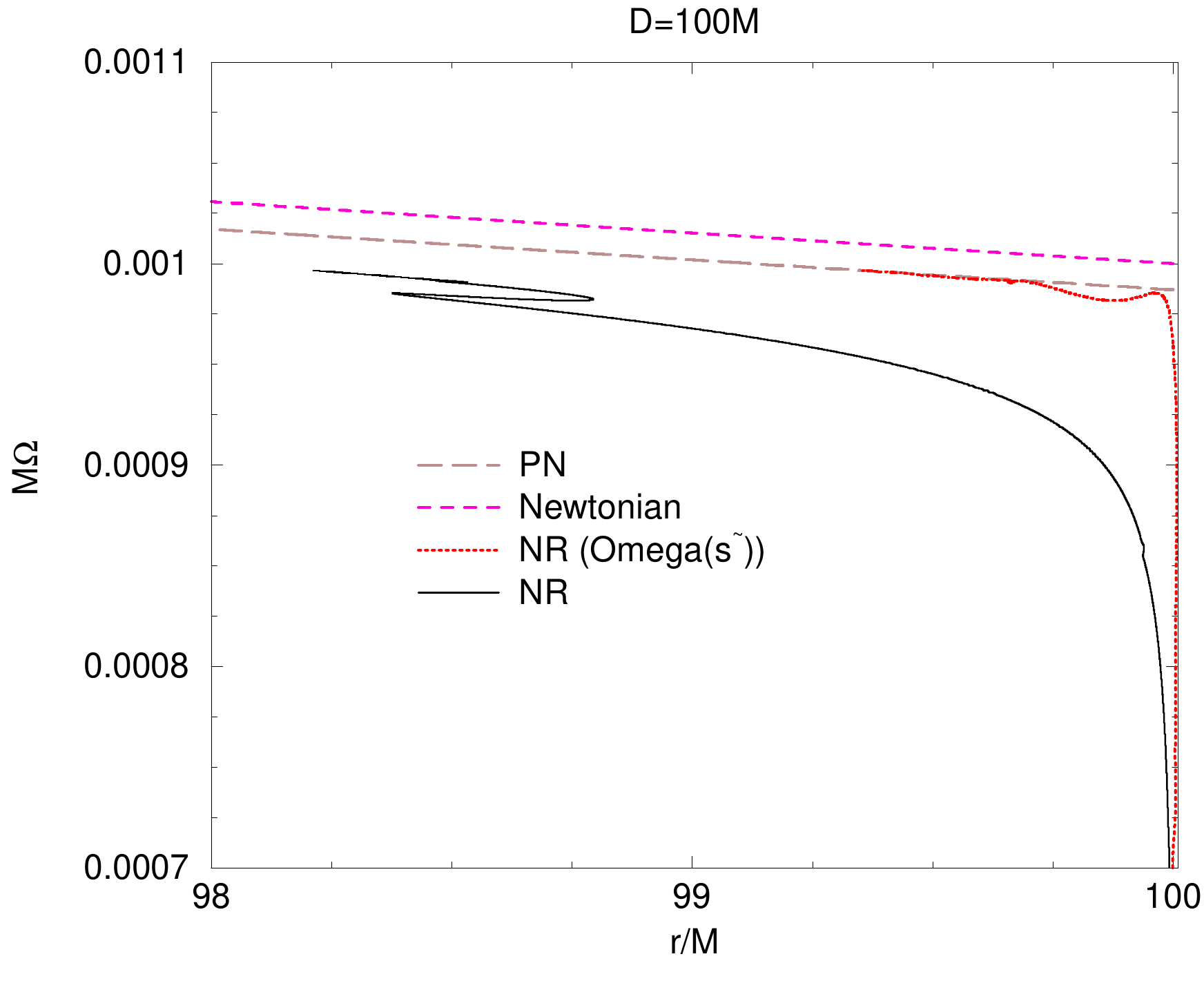}
  \caption{The orbital frequency $d\phi_{\rm orbit}/dt$ versus orbital
separation $r$ and versus {\it shifted semiproper distance} $s\tilde{}$, as well as the Newtonian and 3.5 PN predictions for 
the $D=20M,50M,100M$  \genb runs. Here $s\tilde{} = s - (s_0 - r_0)$, where
$s_0$ is the initial SPD and $r_0$ is the initial PN orbital
separation.}
\label{fig:neon_omega_v_r}
\end{figure}

On of the robust features of both our \gena and \genb runs
is the approach of the full numerical orbital frequency 
to the PN predicted values from below (after the numerical
coordinate {\it settle} into a quasi-stationary regime 
[which takes $\sim 1$ orbit]).
For the $D=20M$ case (see  
Fig.~\ref{fig:neon_omega_v_r}), the numerical and PN predictions 
agree to two significant digits (e.g., $\Omega_{\rm PN} = 0.01142$, while
$\Omega_{\rm num} = 0.01125$, at $D=18.84M$) after the gauge settles down. 
 The progression is similar for the cases of
orbits starting at separations $D=50M$ and $D=100M$, where 
the simulations completed 2.5 and 1.5 orbits, respectively.
The retrograde features in these figures are due to the remaining
(small) eccentricity of the initial quasicircular data and gauge
effects. We report in Table \ref{tab:eccentricity}
the eccentricities of these BHBs, as measured via the formula
$e_s\sim s^2 \ddot s$, where $s(t)$ is the \spd,
as well as the measured orbital period and the 3 PN prediction,
including the decay rate $\Delta r$ after one orbital period.

\begin{table}[t]
\caption{The eccentricity and orbital period of the numerical simulations,
the 3 PN quasicircular orbital periods at the initial separation and
separation after one orbit, and the 3.5 PN orbital inspiral after one orbit.
} \label{tab:eccentricity}
\begin{ruledtabular}
\begin{tabular}{l|llll}
 D/M & $e_s$ & $T_{\rm NR}/M$ & $T_{\rm PN}/M$ & $\Delta r/M$ \\
\hline
20 & 0.0003 & 615 & 601-592 & -0.208 \\
50 & 0.0002 & 2313 & 2283-2279 & -0.054 \\
100 & 0.0006 & 6422 & 6370-6368 & -0.02 \\
\end{tabular}
\end{ruledtabular}
\end{table}

\begin{figure}
  \includegraphics[width=3.15in]{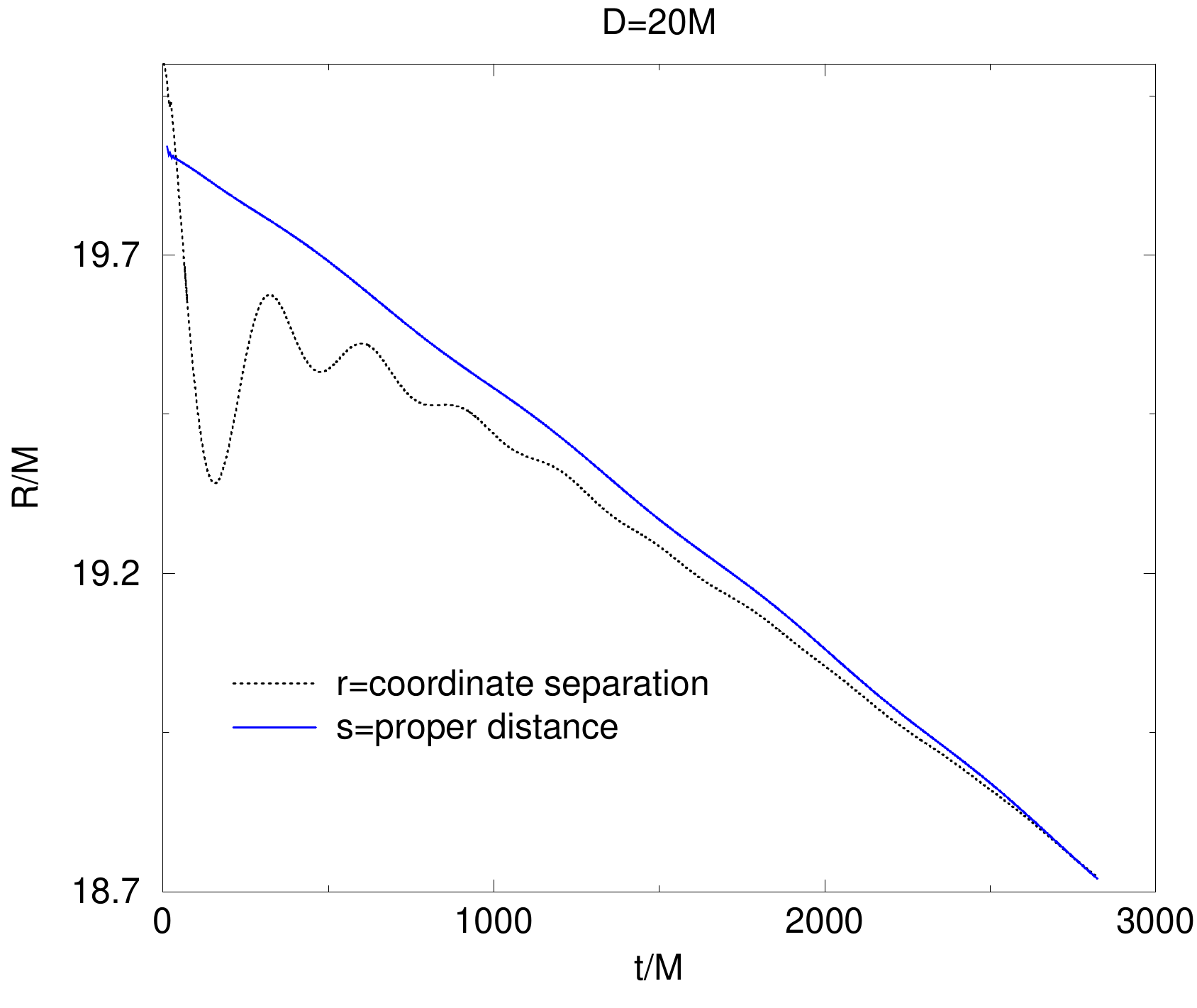}
  \includegraphics[width=3.15in]{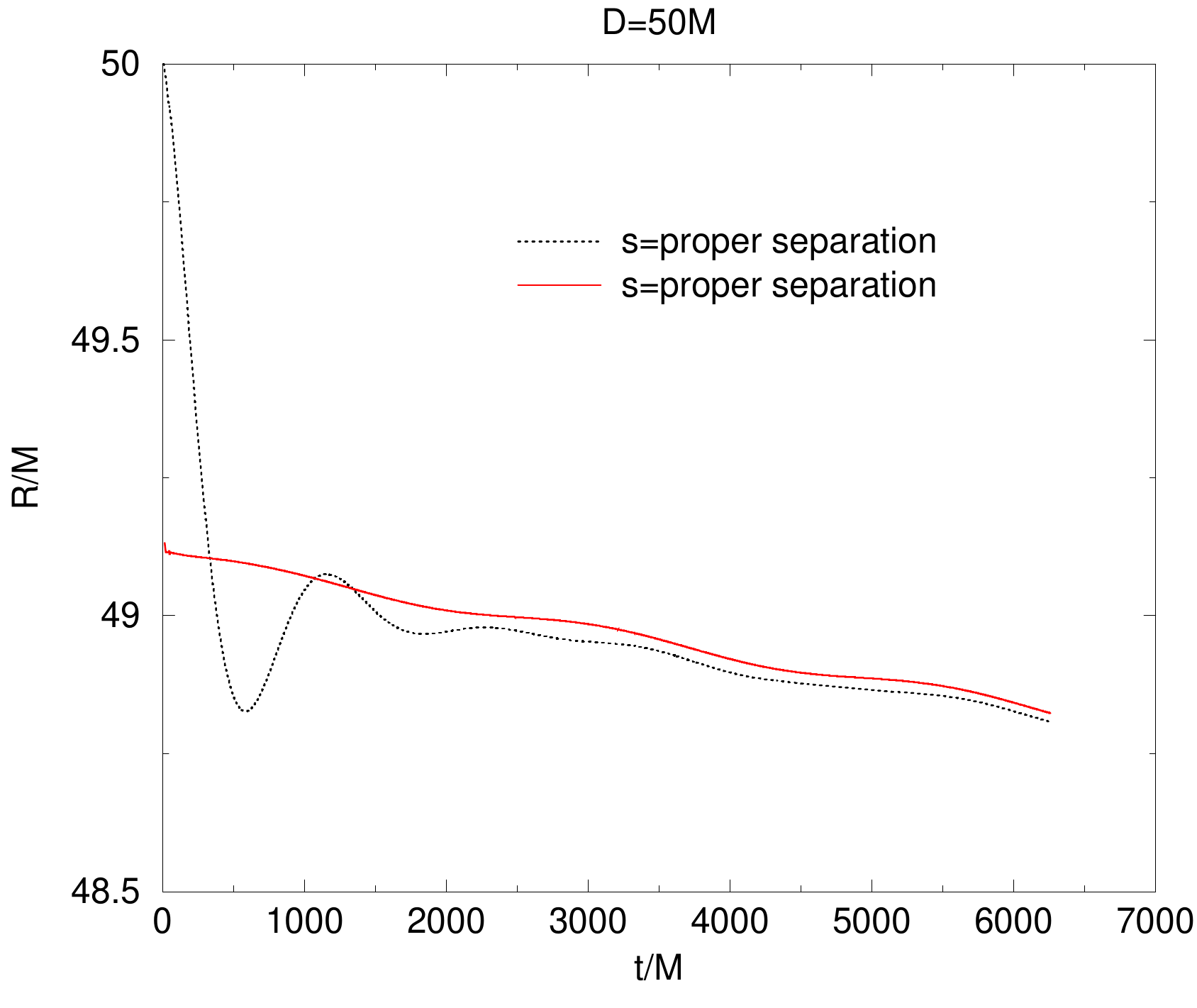}
  \includegraphics[width=3.15in]{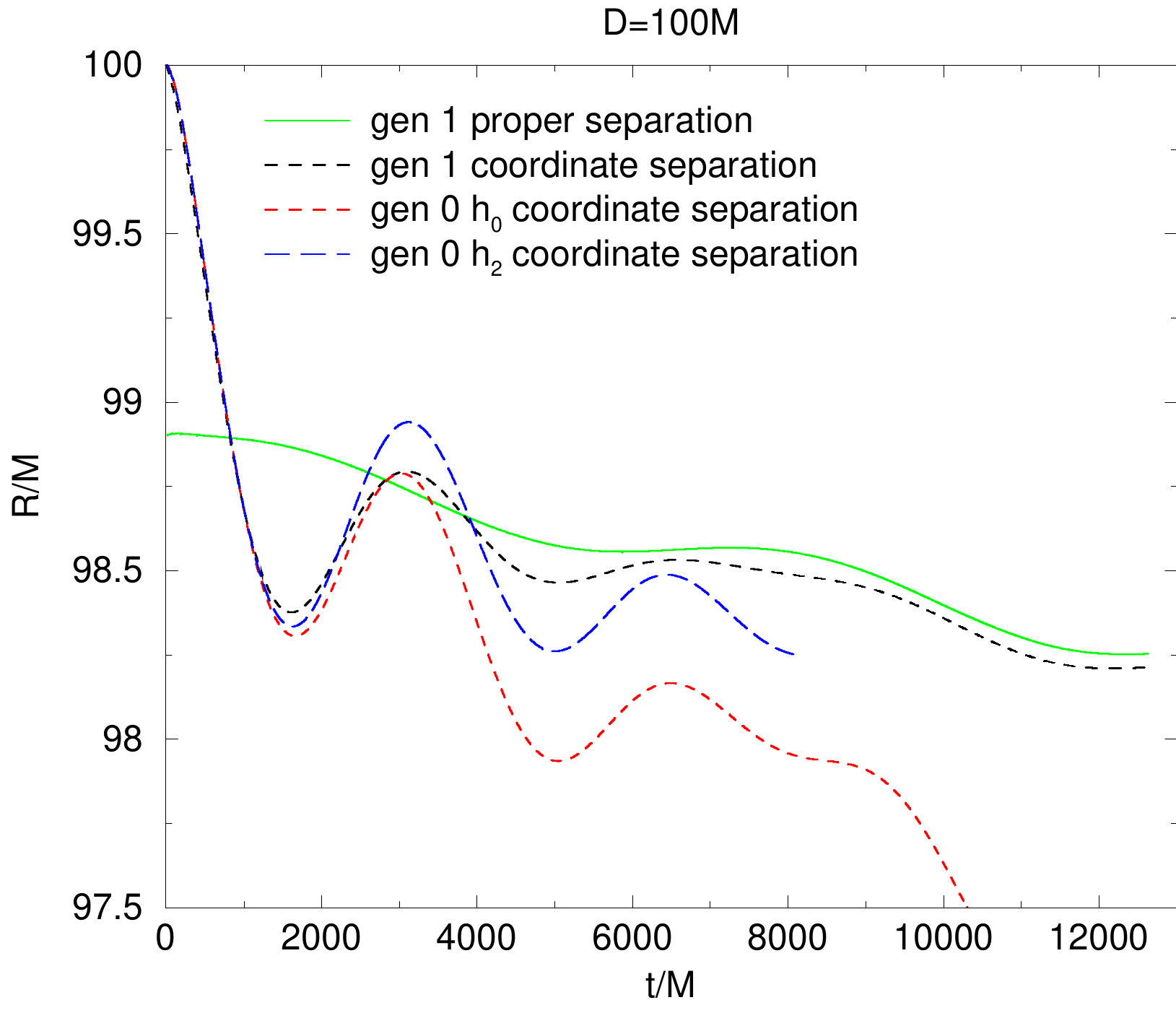}
  \caption{Proper distance (minus a constant) and coordinate distance
comparison for the $D=20M,50M,100M$ \genb runs. The $D=100M$ plot also shows the
coordinate separation versus time for the \gena runs.}
\label{fig:neon_s_r_comp}
\end{figure}

As we saw in the \gena runs (see Fig.~\ref{fig:coord_r}),
the coordinate
separation of the holes shows an interesting gauge effect, namely an
oscillation in time at twice the expected frequency.
In order to show
that this is a non-physical effect, we computed the proper distance
along the line joining the black holes, i.e., the SPD measure
described above (we use this measure since an 
accurate minimal proper distance between horizons is computationally very
intensive). The results are shown in
Fig.~\ref{fig:neon_s_r_comp}. We first observe that the proper distance is
much closer to being monotonic and featureless than the coordinate
separation. We also see that there
are
residual oscillations (of much lower amplitude than those observed in
the coordinate separation) at roughly the orbital frequency, 
which likely are due to the  eccentricity, as well as a possible small residual gauge effect. 

In Fig.~\ref{fig:neon_s_r_comp}, we subtract a constant offset from
the proper distances in order to make comparison with the coordinate
separation easier.
 We see the striking agreement (apart from the overall constant offset
between the two) at
later times, i.e. after the first orbit, showing that the coordinate gauge
effects settle down after one cycle, quite independent from the initial 
radius of the orbit, i.e. in all cases studied, $D/M=20,\,50,\,100$.

\begin{figure}
  \includegraphics[width=3.15in]{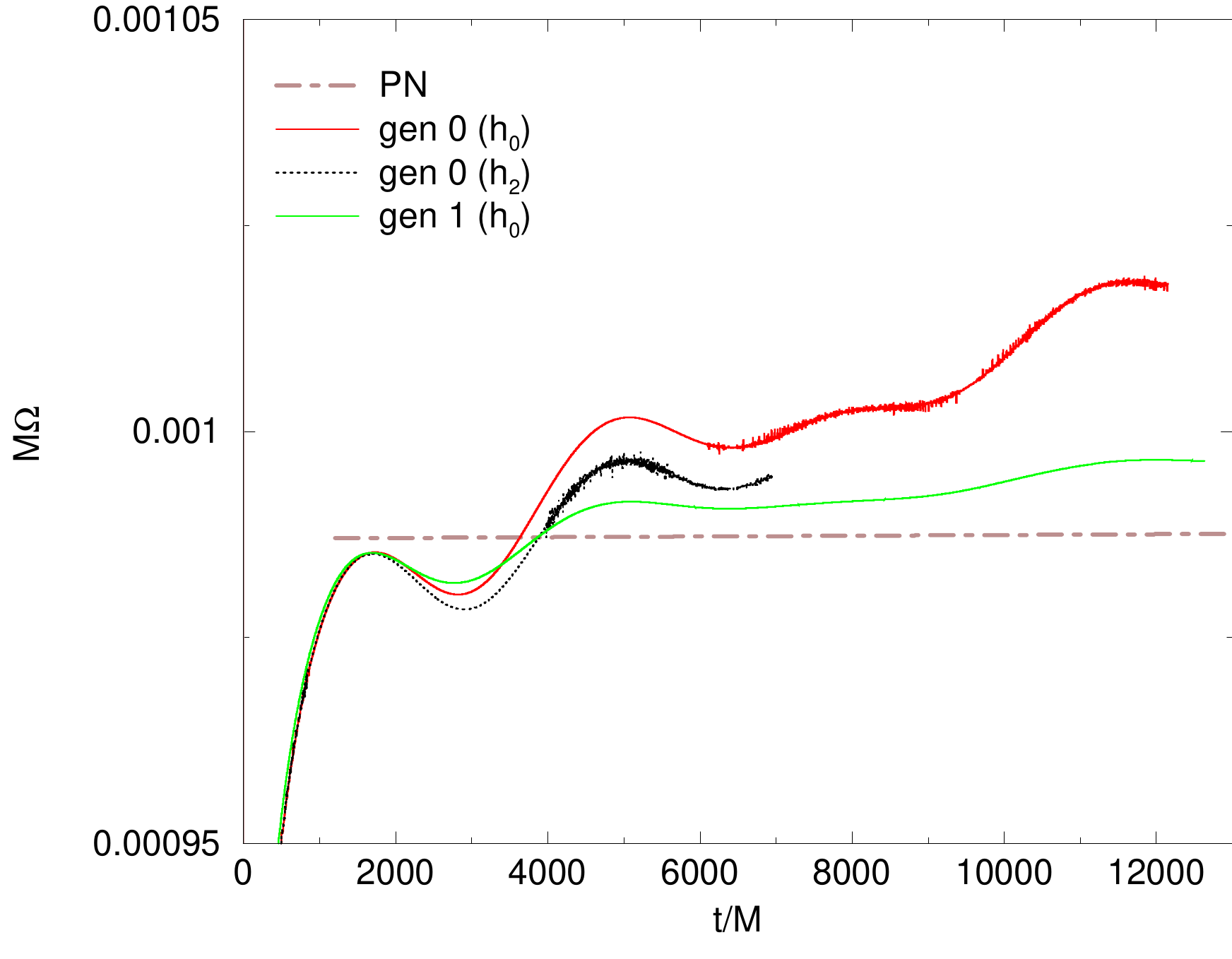}
  \caption{The orbital frequency for the \gena and \genb runs versus
   time ($D=100M$) and the PN prediction. Note how the \genb
appears to outperform the highest resolution \gena run.}
  \label{fig:ru_omega_v_t}
\end{figure}

In Fig.~\ref{fig:ru_omega_v_t}, we compare the orbital frequency, as a
function of time, for the \gena and \genb ($D=100M$) simulations. 
 Note how the \genb
appears to outperform the highest resolution \gena run.


\begin{figure}
  \includegraphics[width=\columnwidth]{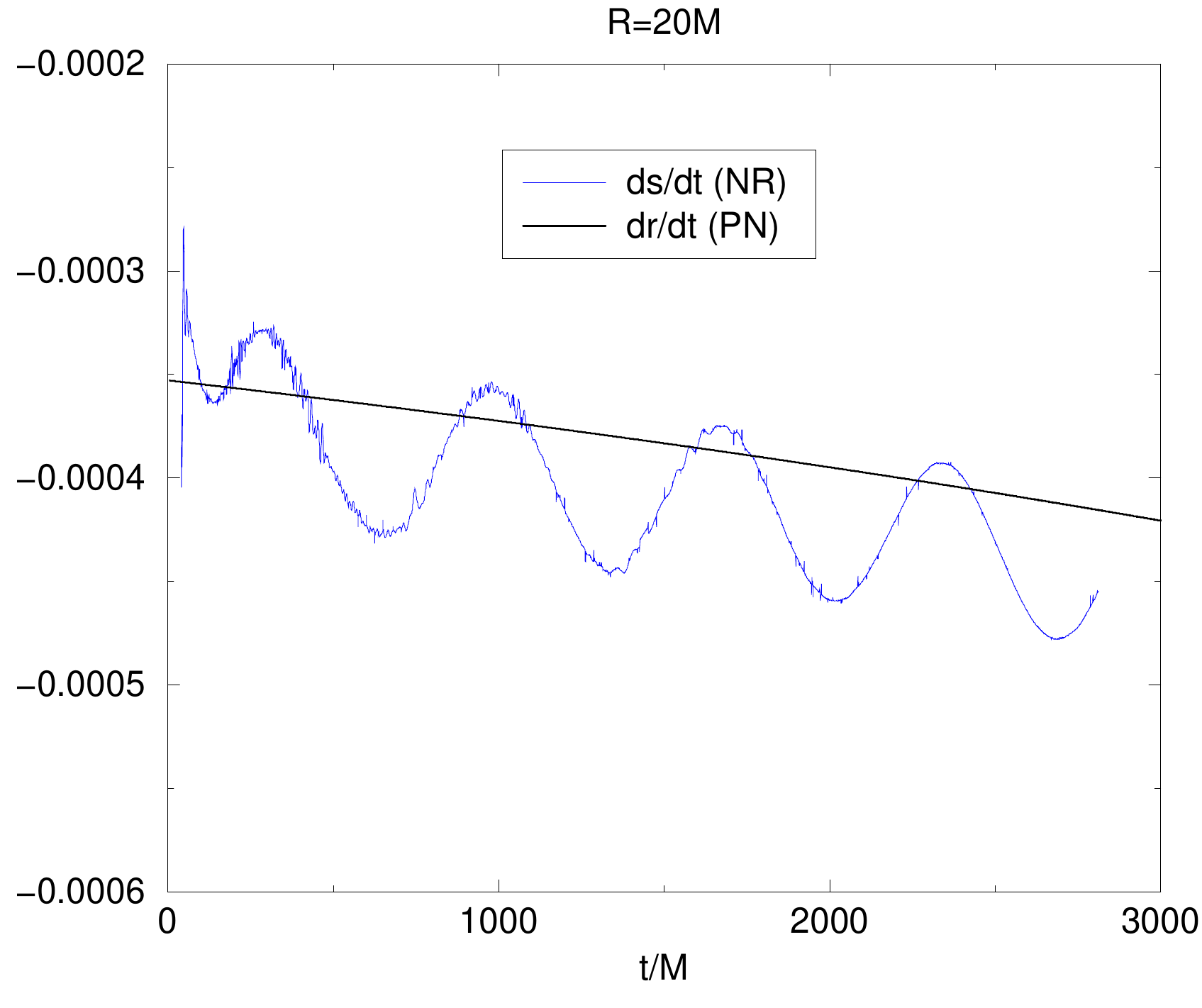}
  \includegraphics[width=\columnwidth]{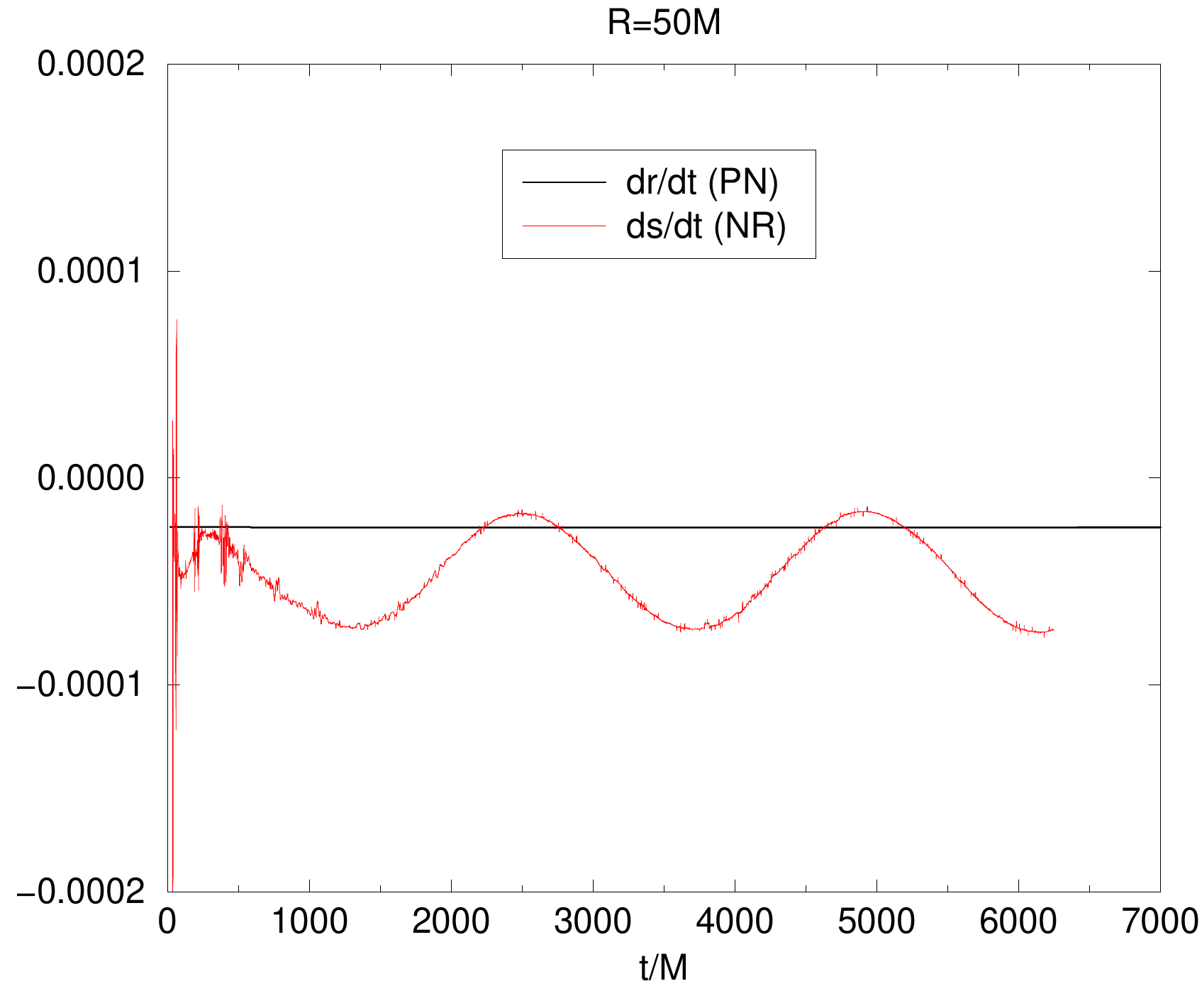}
  \includegraphics[width=\columnwidth]{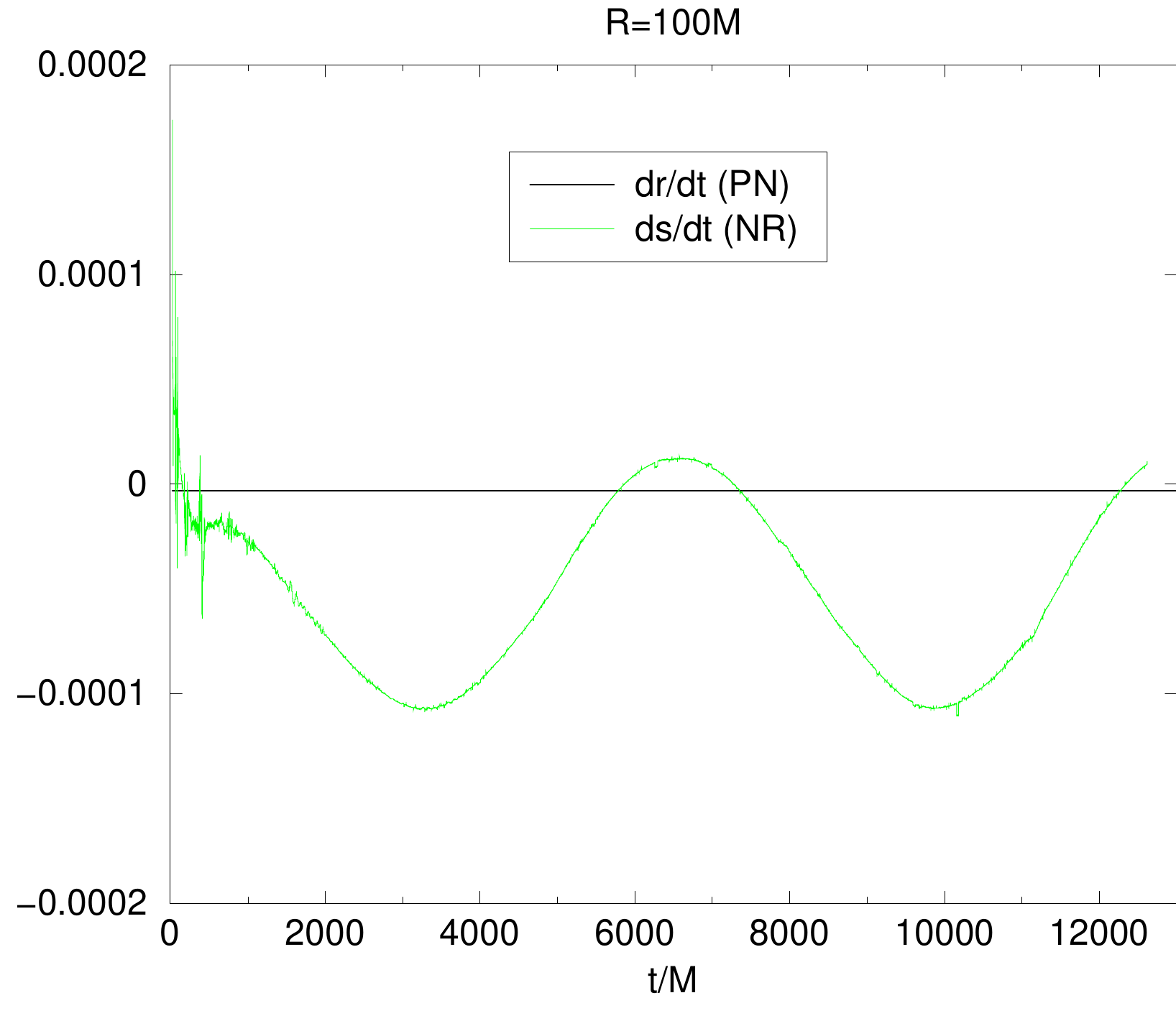}
  \caption{Simple proper distance decay rate versus post-Newtonian prediction
for the $D=20M$,$D=50M$, and $D=100M$ \genb runs. Note that the periods of oscillations roughly match those of the orbital period, $T/M=598,\,2281,\,6369$ respectively}
\label{fig:neon_pn_sdot_cmp}
\end{figure}

Using our more accurate \genb simulations, and the SPD measure of
the orbital separation,
we study in more detail the orbital
decay of the numerical runs  and
compare them with the PN predictions.
Here were examine $\dot s(t)$ as a function of time, where $s(t)$
is the SPD.
 In Fig.~\ref{fig:neon_pn_sdot_cmp}
we see the good agreement between the mean values of $\dot s(t)$ and
the 3.5PN predictions. The agreement is slightly better for the closer 
$D/M=20$  case, which we evolved for three orbits,
 than for the more demanding $D/M=50,\,100$ cases, which
did not complete two orbits. 
There is an oscillation
corresponding closely to the orbital period of the $D/M=20,\,50,\,100$ 
runs (according to PN, $T/M=598,\,2281,\,6369$, respectively). Note that in 
Fig.~\ref{fig:neon_pn_sdot_cmp} we used the same ordinate scale and
therefore
the amplitudes of oscillations are comparable, which indicates
that the eccentricities of the farther separated BHBs are larger.

\begin{figure}
  \includegraphics[width=3.0in]{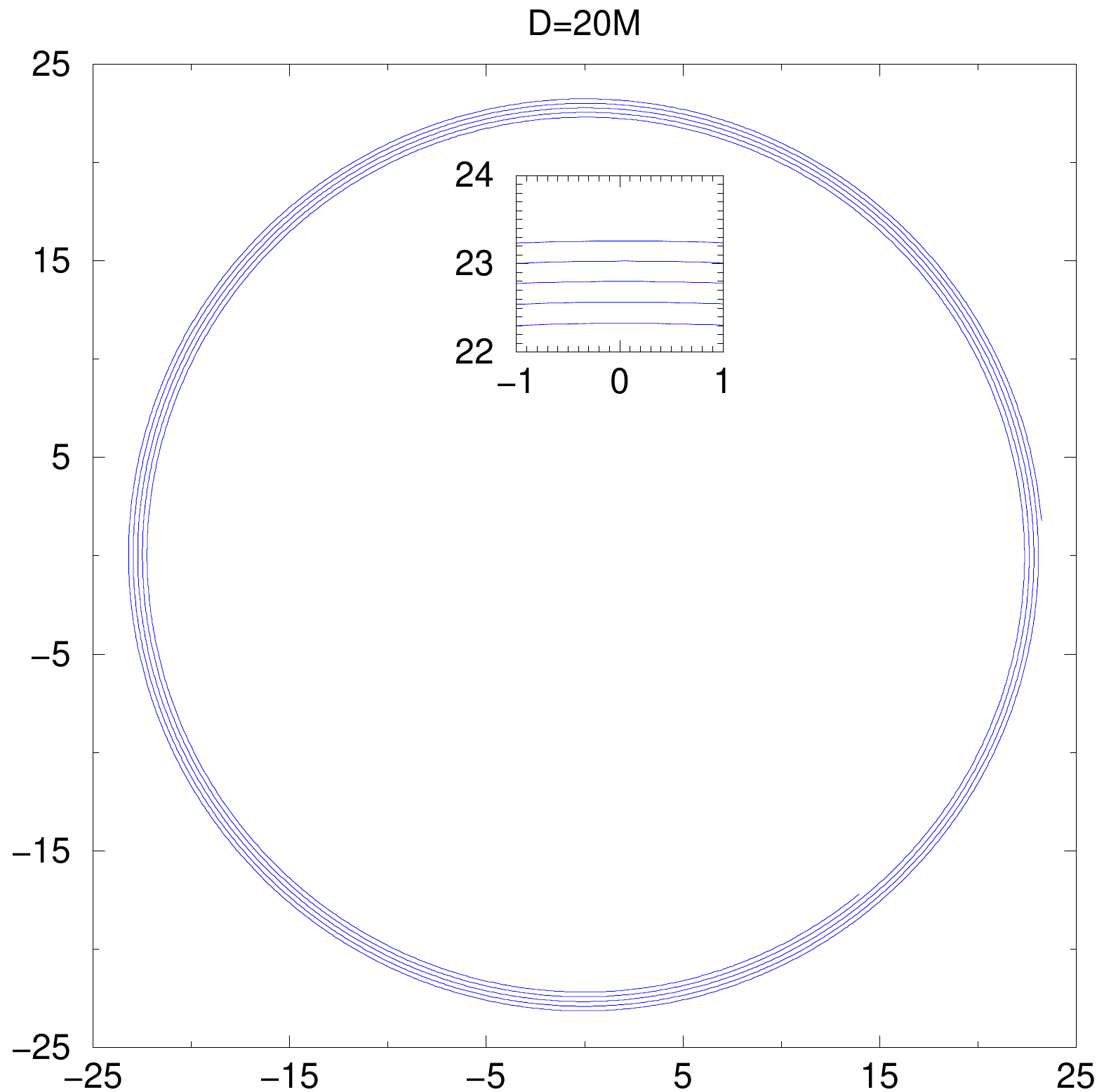}
  \includegraphics[width=3.0in]{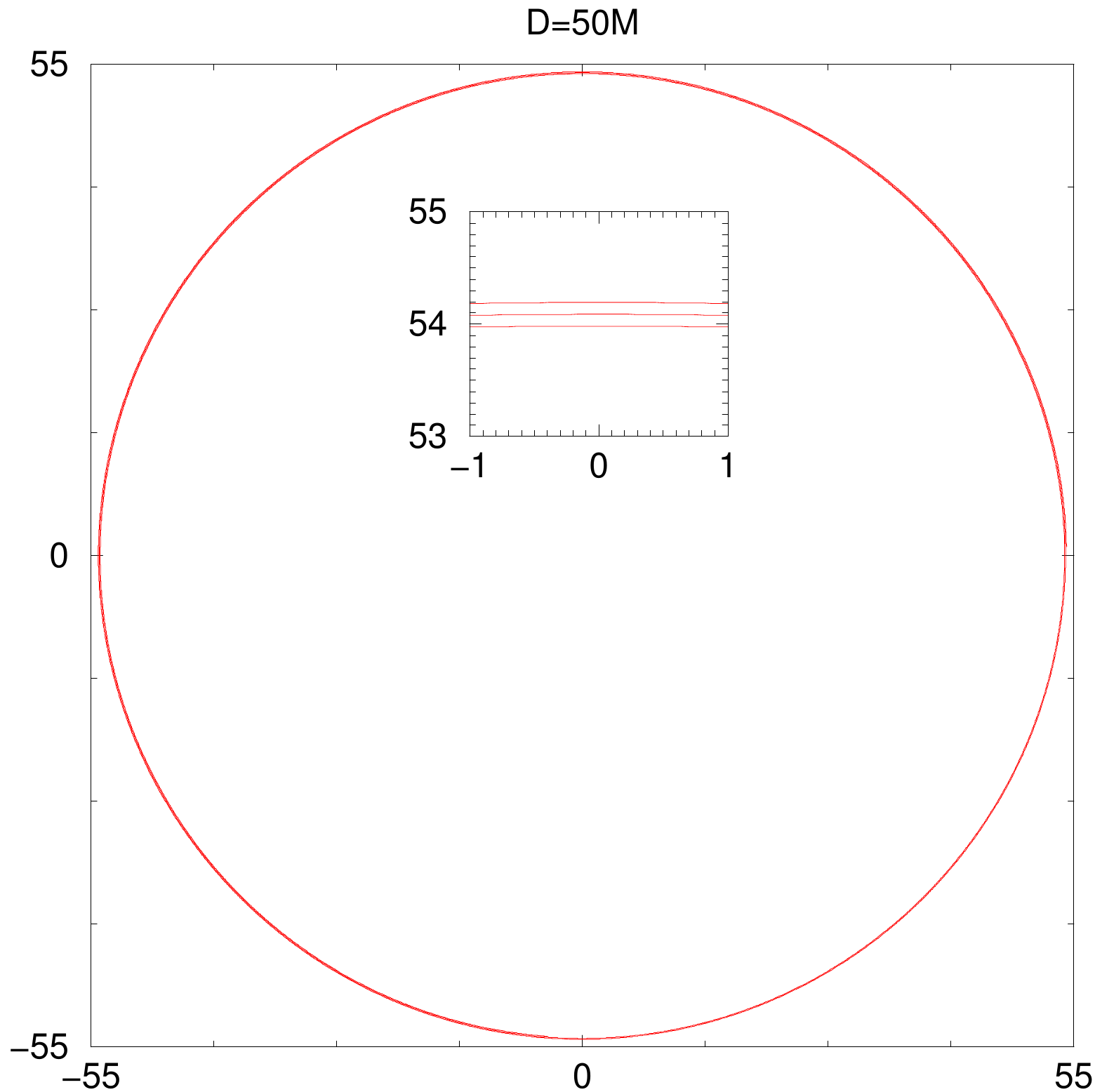}
  \includegraphics[width=3.0in]{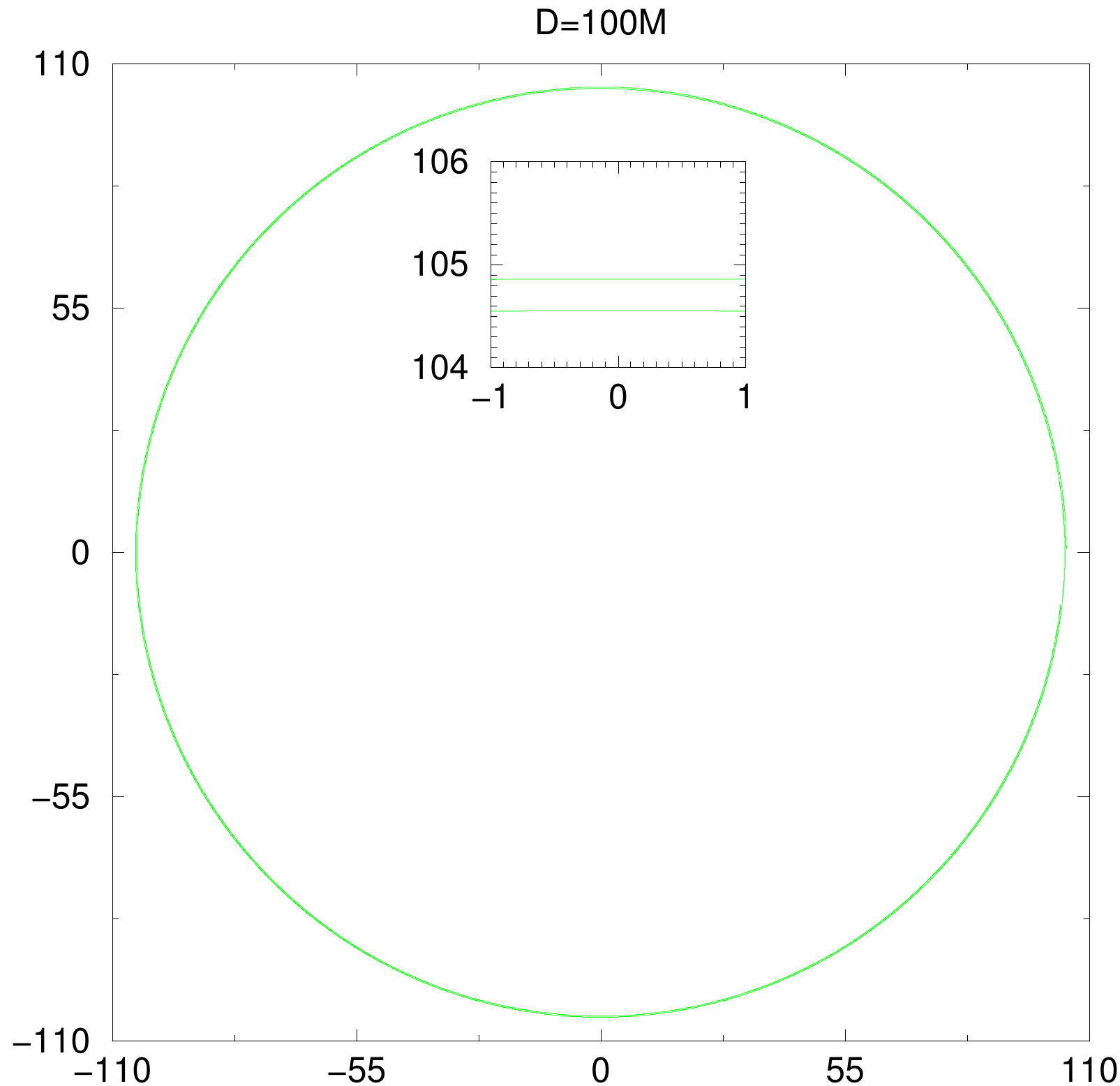}
  \caption{The orbital {\it trajectory} obtained using the coordinate
orbital phase and the {\it proper} orbital separation along the
coordinate line separating the two BHs for the \genb simulations.
 The zoomed plots show  the tightness of the spiral. }
\label{fig:semiproper_orbit}
\end{figure}

In order to have a visual notion of how tight the orbits look in
proper distance $s$ space, we plot the {\it semiproper} trajectory
$\vec r(\phi) = s (\cos\phi,\sin\phi)$. Here $\phi$ is the coordinate
azimuthal angle, while $s$ is the SPD. Our results are shown
in Fig.~\ref{fig:semiproper_orbit}. The scale of the zoom-in plots
is the same for all runs (making
comparison of the relative tightness of the orbits
straightforward). Note that the
 larger separation between successive orbits for the  $D=100M$ case
than for the $D=50M$ case  likely reflects some combination of residual gauge
effects and truncation errors (see Fig.~\ref{fig:neon_pn_sdot_cmp}).
While the $D=100M$ orbital decay rate appears to be too large by a factor of
2-3, the gauge invariant radiated energy is consistent with 
PN predictions (see Sec.~\ref{sec:pn-nr-wave} and
Fig.~\ref{fig:neon_pn_sdot_cmp}).

We note that, according to 3.5PN evolutions, the $D=20M$ BHB completes 
36 orbits and takes $t=14,200M$ to inspiral to $D=5M$. The $D=50M$
BHB, on the other hand completes 370 orbits and takes $t=531,716M$ to
get to the same separation. Finally, the $D=100M$ BHB completes 2064
orbits and takes $t=8,223,170M$.

\subsection{Comparison with Post-Newtonian waveforms}
\label{sec:pn-nr-wave}

The leading $(2,2)$-mode of the PN waveform~\cite{Faye:2012we} 
is given by
\beq\label{eq:h22}
r_{\rm obs}\ h^{(2,2)}=-8m\sqrt{\frac{\pi}{5}}\eta\,x\,H^{(2,2)}e^{-2i\psi}
\eeq
where 
$\psi=\varphi-2M\Omega\ln(\Omega/\Omega_0)$, $\varphi$ is the orbital
phase, 
\bea\label{eq:H22}
H^{(2,2)}&=&1+\left(-\frac{107}{42}+\frac{55}{42}\eta\right)\,x+2\pi\,x^{3/2}\nonumber\\
&&+\left(-\frac{2173}{1512}-\frac{1069}{216}\eta+\frac{2047}{1512}\eta^2\right)\,x^2\\
&&+\left(-\frac{107}{21}\pi+\frac{34}{21}\pi\eta-24i\eta\right)\,x^{5/2}+\cdots\nonumber
\eea
$x=(m\Omega)^{2/3}$, and $\Omega_0$ is an arbitrary positive
constant (see \cite{Faye:2012we})
which we took to be the initial orbital frequency.
 For quasicircular orbits, up to 3 PN, the orbital
frequency is given by
\footnote{Provided by H. Nakano based on \cite{Schafer:2009dq}}
\bea
m\Omega&=&\left(\frac{m}{R}\right)^{3/2}
\left[1+\frac{(\eta-3)}{2}\left(\frac{m}{R}\right)\right.\nonumber\\
&&\left.+\frac{(6\eta^2+7\eta+24)}{16}\left(\frac{m}{R}\right)^2
\right.\nonumber\\
&&\left.+\frac{(120\eta^3-612\eta^2+501\pi^2\eta-8096\eta-480)}{384}
\left(\frac{m}{R}\right)^3\right.\nonumber\\
&&\left.+\cdots\right],
\eea
where $R$ denotes the ADM-TT radial coordinate.

For large $r$, $\psi_4$ is given by~\cite{Fujita:2010xj}
\beq\label{eq:PNWaveform}
\psi_4=\frac12\left(\ddot{h}_+-i\ddot{h}_\times\right)
\approx{-2m^2\Omega^2}h^{(2,2)}.
\eeq

The radiated energy and angular momentum are give by Eqs. (7.4a)-(7.4b) of Ref.~\cite{Arun:2009mc}
\begin{subequations}\begin{align}
\frac{dE}{dt}&= \frac{32
}{5}\,\eta^2\,x^5\,\left\{1-\left(\frac{1247}{336}+ \frac{35}{12}\eta
\right)x+4\,\pi\, x^{3/2}\right.\nonumber\\
&\left.+\left(-\frac{44711}{9072} + \frac{9271}{504} \eta + \frac{65}{18} \eta
^2\right)x^2\right.\nonumber\\
&\left.+\left(-\frac{8191}{672}-\frac{583}{24}\eta\right)x^{5/2}
\right\}+\cdots\\
\frac{dJ}{dt}&=\frac{32}{5}\,m \,\eta ^2 \,x^{7/2} \,\left\{1-\left(\frac{1247}{336}
+ \frac{35}{12}\eta \right)x\right.\nonumber\\
&\left.+ 4\pi\,x^{3/2}+\left(-\frac{44\,711}{9072} + \frac{9271}{504} \eta +
\frac{65}{18} \eta ^2\right)x^2\right.\nonumber\\
&\left.+\left(-\frac{8191}{672}-\frac{583}{24}\eta\right)x^{5/2}
\right\}+\cdots
\end{align}\end{subequations}
where $\frac{dE}{dt}=\Omega\frac{dJ}{dt}$ for quasicircular orbits.
Hence, we can approximate the energy $E$ and angular momentum $J$ radiated per 
quasicircular orbit by
\beq
\delta E=\int_0^T
\frac{dE}{dt}dt\approx\frac{dE}{dt}T=\frac{dE}{dt}\frac{2\pi}{\Omega},
\eeq
with a similar expression for the radiated angular momentum (where the
integral is over the time interval corresponding to two complete
cycles of the $(\ell=2, m=2)$ mode of $\psi_4$, and hence one orbit
of the BHB).

\begin{figure}
  \includegraphics[width=\columnwidth]{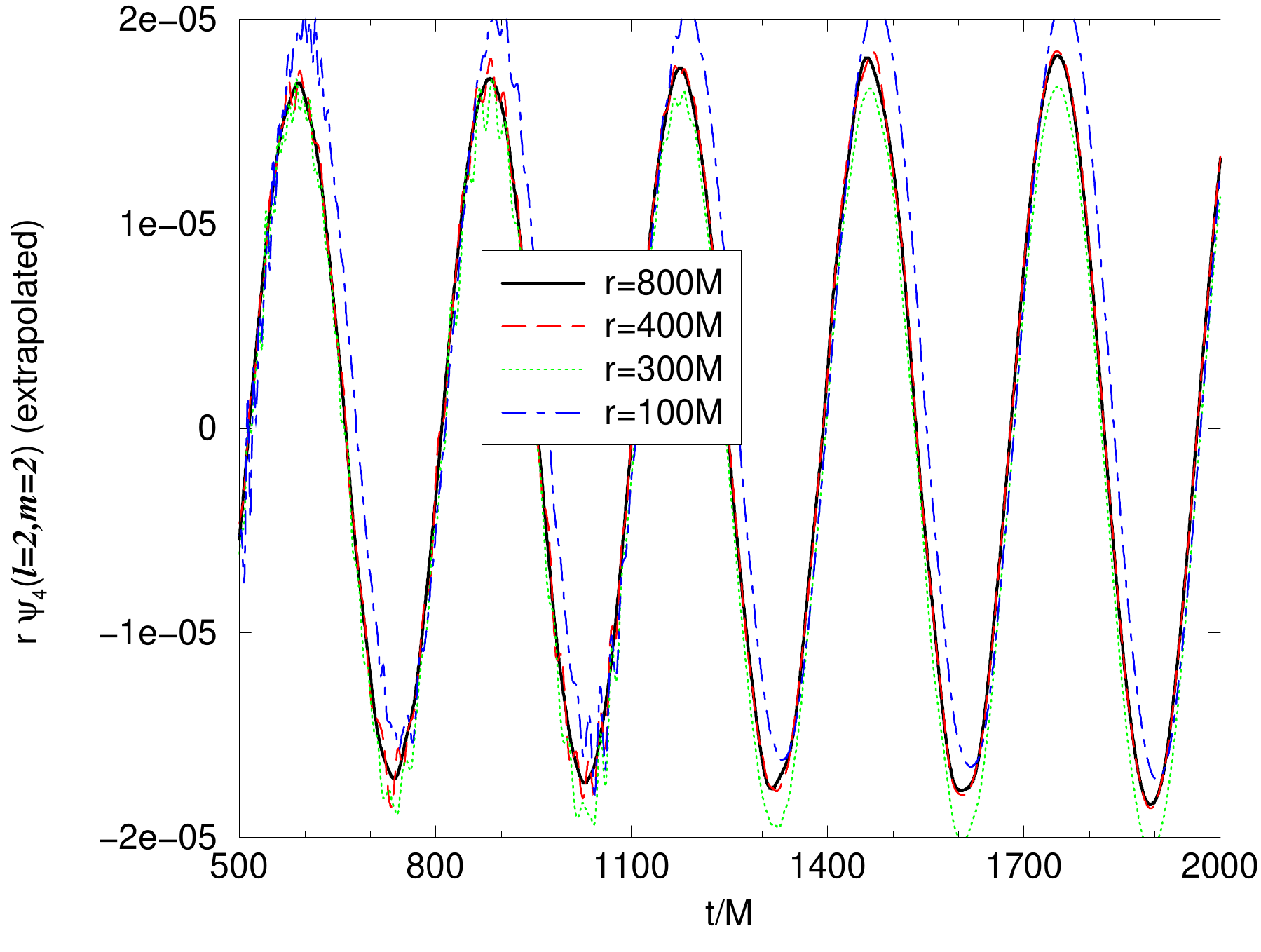}
  \caption{The waveform extracted at different observer locations
for the $D=20M$ \genb run. Here the waveforms have been {\it extrapolated to
$\infty$} using Eq.~(\ref{eq:asymtpsi4ext}). 
While we see that the extraction at $R_{\rm obs}=100$ has
large phase errors, the phases and amplitudes stabilize
for $R_{\rm obs}\geq300$, i.e. one wavelength distance from the sources.}
\label{fig:neon_wave}
\end{figure}

\begin{figure}
  \includegraphics[width=\columnwidth]{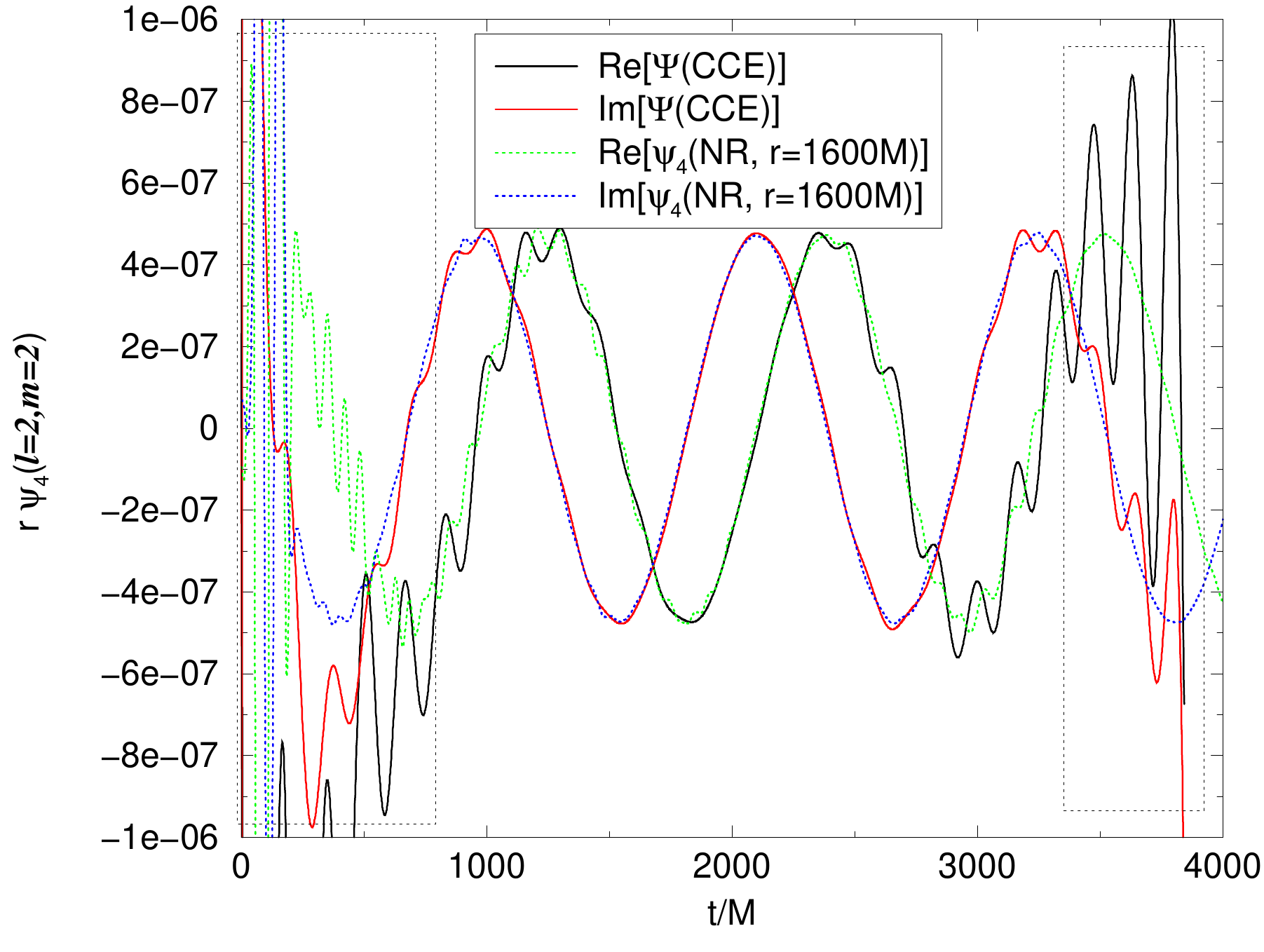}
  \caption{Comparison of $r\psi_4$ and the Null variable $\Psi$ for
the
$D=50M$ simulation. The boxes indicate regions where the fast Fourier transform smoothing
operation on the CCE data distorted the waveform.}
\label{fig:tin_cce_nr_wave}
\end{figure}

Comparisons between full numerical and PN waveforms can 
provide a strong test of both approaches.
For the \genb runs,
 we extended the computational
domain to a cube of sides $6400M$ in order to be able to extract
the gravitational waveforms 
in the {\it radiation} zone.
The effects of the extraction radii on the
amplitude and phase of the numerical waveforms is apparent in 
Fig.~\ref{fig:neon_wave}, where we plot
the $D=20M$ the waveforms as seen by observers at
$R_{\rm obs}/M=100,\,300,\,400,\,800$. Note that in this case, the
half orbital period (measuring the period of the waves) is
$\approx300M$. We see that extracting at $R_{\rm obs}/M=100$ leads to 
inaccuracies in both the amplitude and phase, but for
$R_{\rm obs}/M=300,\,400,\,800$ the waveforms line up; indicating that
extracting at least at one wavelength from the system is necessary.
In all cases, we used the perturbative extrapolation
formula given in Ref.~\cite{Lousto:2010tb}.
\begin{eqnarray}
&&\lim_{r\to\infty}[r \,\psi_{4}^{\ell m}(r,t)]  \nonumber \\
&&=\left[r \,\psi_{4}^{\ell m}(r,t) 
- \frac{(\ell -1)(\ell +2)}{2} \int_0^t dt \, \psi_{4}^{\ell m}(r,t)
\right]_{r=R_{\rm obs}}\nonumber\\
&&+ O(R_{\rm obs}^{-2}) \,,
\label{eq:asymtpsi4ext}
\end{eqnarray}
where for our choice to the tetrad we have
$\psi_4=(1/2-M/r)\,\psi_4^{\rm Num}$.
For larger orbital separations the orbital period increases
(See Eqs.~(\ref{eq:Omega}-\ref{eq:T}))
approximately as $\sim D^{3/2}$ hence we need observer
locations of at least $R_{\rm obs}/M=1150$ and $R_{\rm obs}/M=3200$,
respectively, for binaries with separations $D/M=50$ and $D/M=100$.
In Fig.~\ref{fig:tin_cce_nr_wave}, as an independent 
validation of the finite 
radius extraction,
we compare the perturbative
extraction as defined above (\ref{eq:asymtpsi4ext}) with the
Cauchy characteristic extraction (CCE) code described in 
Ref.~\cite{Babiuc:2010ze}
and observe the good agreement among them in the common region of
validity.

\begin{figure}
  \includegraphics[width=\columnwidth]{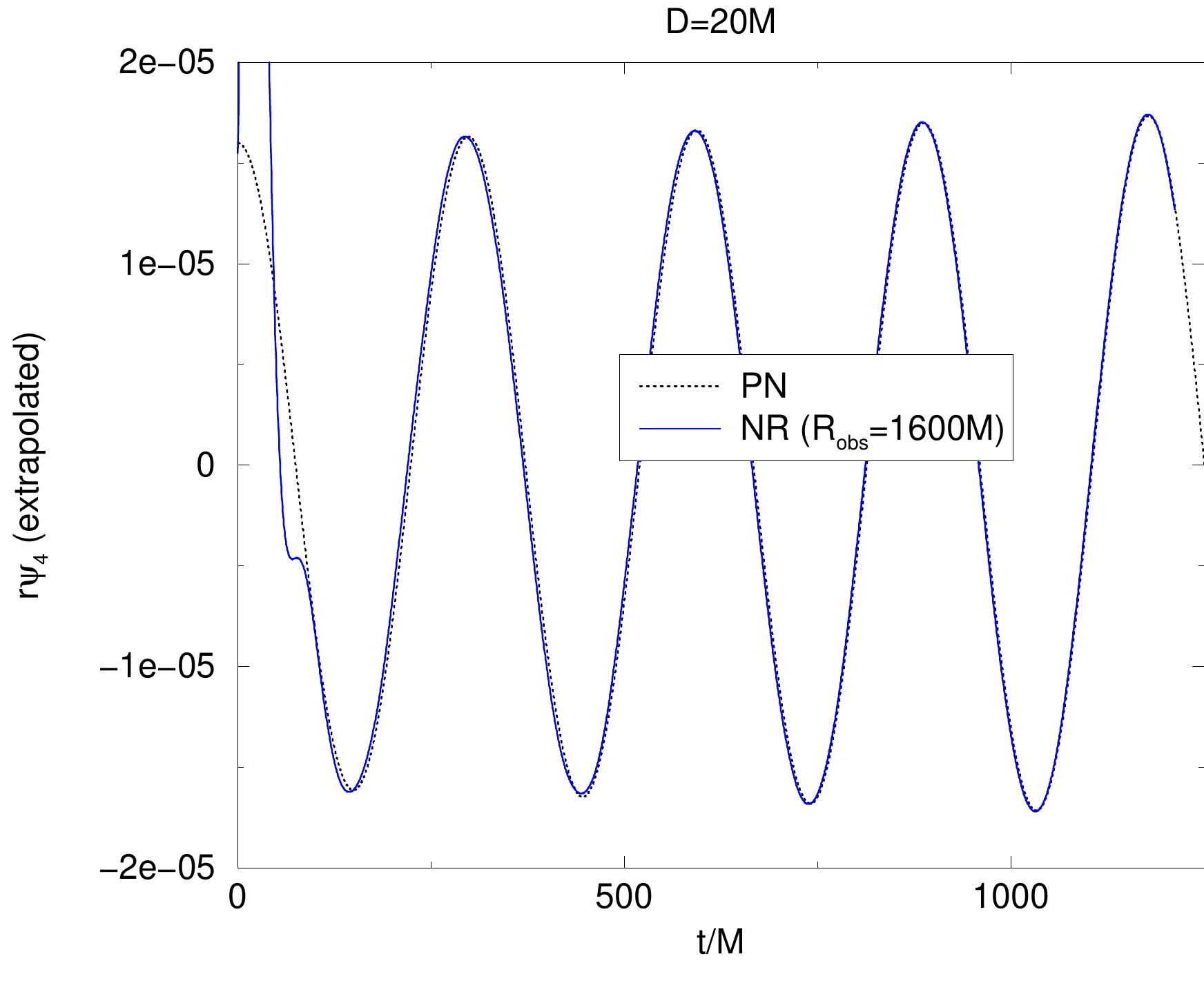}
  \includegraphics[width=\columnwidth]{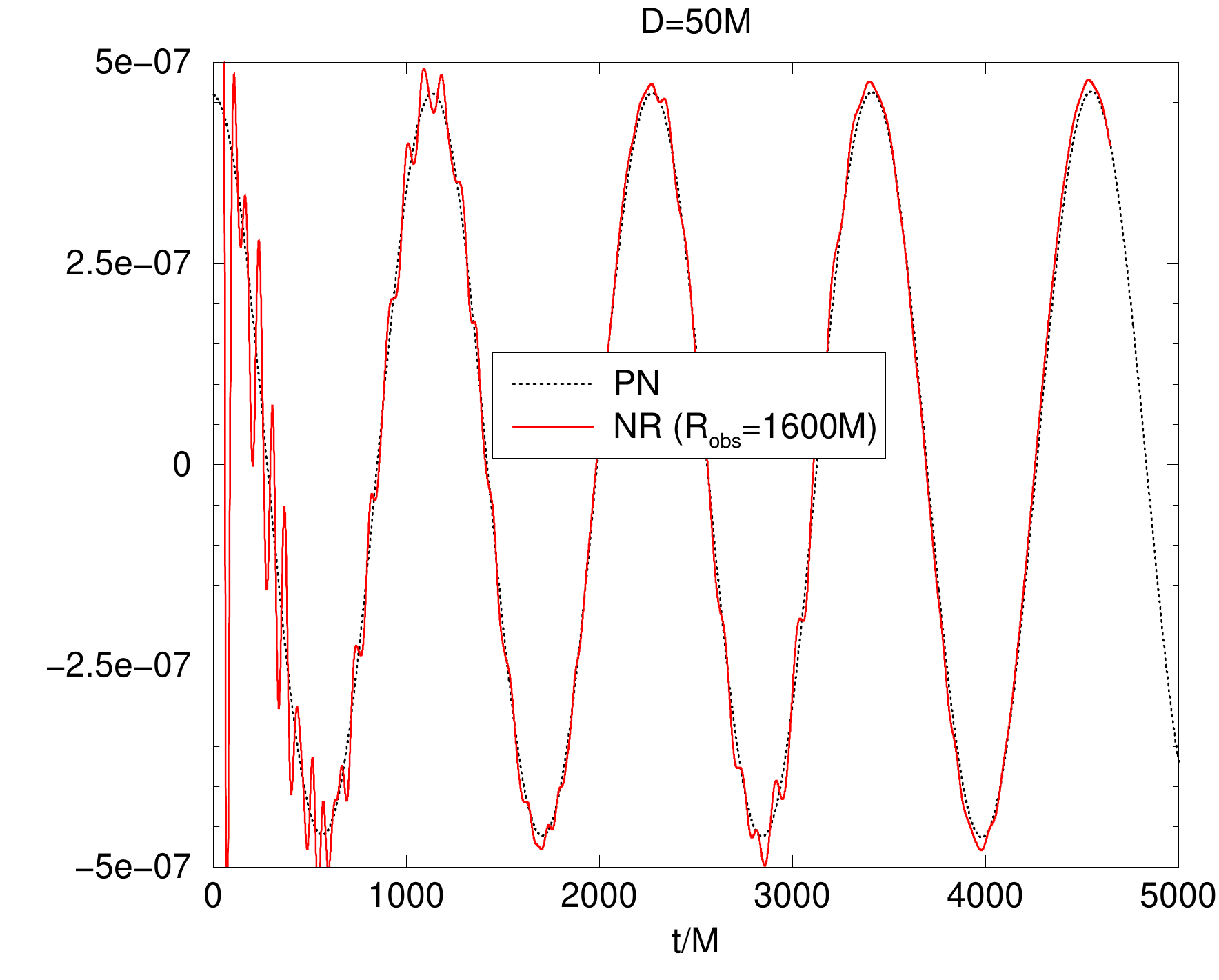}
  \includegraphics[width=\columnwidth]{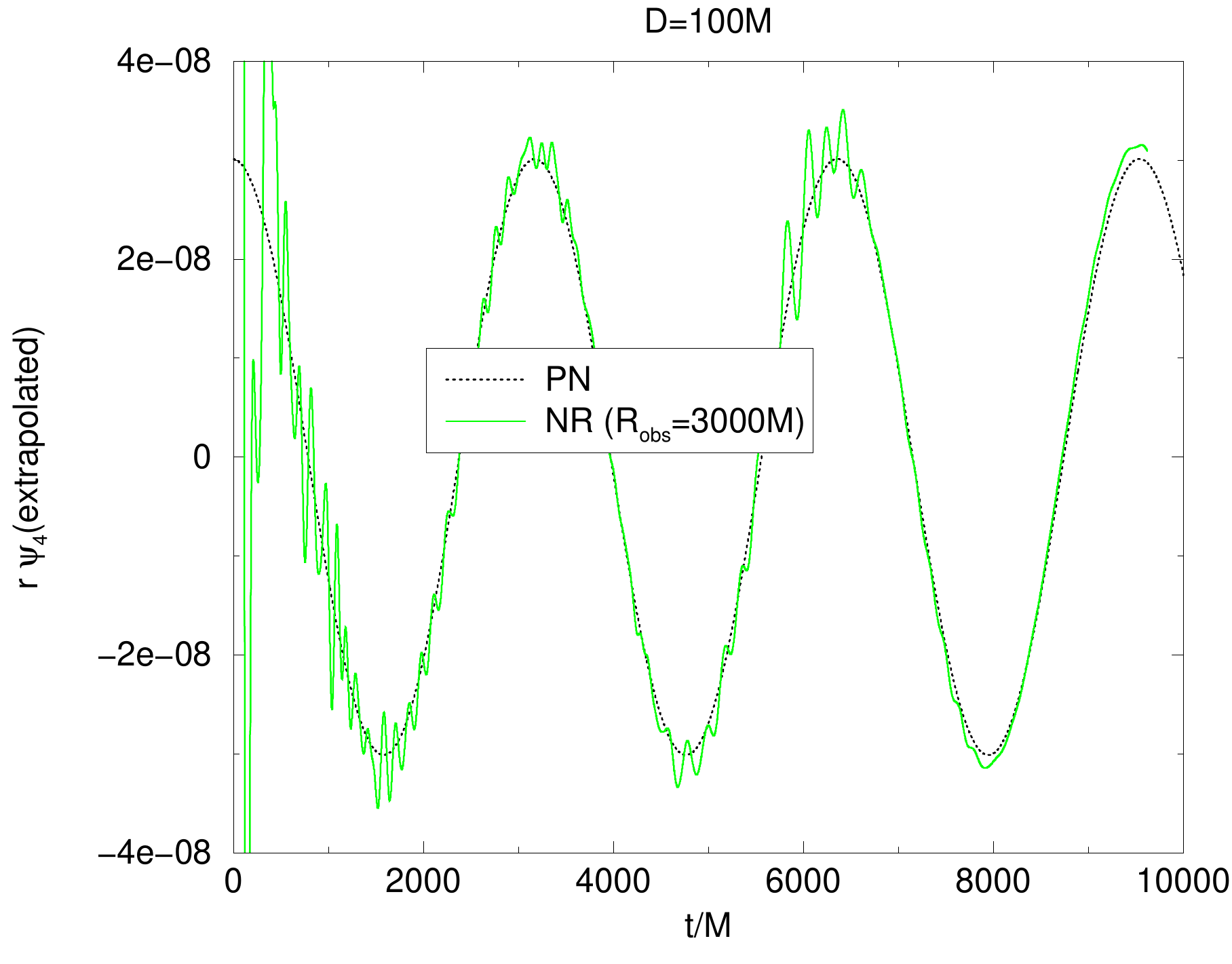}
  \caption{Comparison of 3.5PN and full numerical waveforms for the
$D=20M,50M,100M$  \genb simulations. Note the excellent agreement in both 
amplitude and phase and the very small scale of the amplitudes.}
\label{fig:ne_wave_pn_comp}
\end{figure}

We find a striking agreement (particularly at larger separations)
 between these full numerical waveforms 
and the PN waveforms
given by Eq.~(\ref{eq:PNWaveform}). 
In Fig.~\ref{fig:ne_wave_pn_comp}
we plot $R \psi_4$ extrapolated using
Eq.~(\ref{eq:asymtpsi4ext}) and the 3.5PN~\cite{Fujita:2010xj}
prediction. The differences in
the phase and amplitudes are within the numerical noise.
For the $D=20$ case, we still observe improvements between the 2PN expression,
as truncated in Eq.~(\ref{eq:H22}), and the 3.5PN expression.
 At the larger initial separations,
$D=50M,100M$, even the lower-order PN expressions [i.e.\
Eq.~(\ref{eq:H22})]
show excellent agreement with the full numerical waveform.

We note that based on the  \gena convergence simulations, we
may expect that the orbital phase error in the $D=100M$ simulation run is
under $0.025$ rad (the argument being that the \genb simulations should
be more accurate at a given resolution than the \gena simulations).
Based on the observed agreement between the PN and NR simulation, it
appears that the phase error is indeed (relatively) small.

Finally, we give the energy and angular momentum radiated, per orbit, in
Table~\ref{tab:rad}. For the table, we measure the energy and
angular momentum radiated over one period of the $(\ell=2,m=2)$ mode
(however, we use all modes up to $\ell=4$ to calculate the energy and
angular momentum radiated), we then multiply by 2 (we do this because
we do not have two full cycles at $R_{\rm obs}=3000M$ for $D=100M$).
 We approximate the
error in these measurements by calculating $\delta m$ and
$\delta J$ at $R_{\rm obs}=3000M$ ($R_{\rm obs}=1600M$ for $D=20M$)
and $R_{\rm obs}=1500M$ ($R_{\rm obs}=800M$ for $D=20M$) and
take the
difference between these two measurements as the error.
We compute the energy and angular momentum radiated at the initial
separation and after one orbit including the radial decay given by
$\Delta r\approx\langle\frac{\ud a_r}{\ud t}\rangle\,T,$ where
\cite{Arun:2009mc}
\bea
\left\langle\frac{\ud a_r}{\ud t}\right\rangle &=&
-\frac{64}{5} c x^3 \eta\left\{1+\left({-\frac{743}{336} -\frac{11}{4} \eta
}\right)x+4\pi\,x^{3/2}\right.\nonumber\\
&&\left.+\left(\frac{38\,639}{18\,144}+ \frac{11\,393}{2016} \eta + \frac{19}{6} \eta^2\right)x^2\right.\nonumber\\
&&\left.+\pi\left(-{4159\over672}-{189\over8}\,\eta\right)x^{5/2}+\cdots\right\}.
\eea

\begin{table}[t]
\caption{Energy and angular momentum radiated per
orbit (initial and second).} \label{tab:rad}
\begin{ruledtabular}
\begin{tabular}{l|ll}
 D & $\delta m/M_{\rm num}$ & $\delta m/M_{\rm PN}$ \\
\hline
20M & $(5.68\pm0.02)\times10^{-5}$ & $5.43-5.62\times10^{-5}$ \\
50M & $(2.4\pm0.1)\times10^{-6}$ & $2.52-2.53\times10^{-6}$  \\
100M & $(2.3\pm0.4)\times10^{-7}$ & $ 2.36-2.36\times10^{-7}$ \\
\hline
\hline
 D  & $\delta J/M^2_{\rm num}$ & $\delta J/M^2_{\rm PN}$\\
\hline
20M & $(5.39\pm0.01)\times10^{-3}$ & $5.20-5.30\times10^{-3}$\\
50M & $(8.9\pm0.1)\times10^{-4}$ & $9.16-9.18\times10^{-4}$\\
100M &  $(2.4\pm0.4)\times10^{-4}$ & $2.39-2.39\times 10^{-4}$\\
\end{tabular}
\end{ruledtabular}
\end{table}

\section{discussion}\label{sec:discussion}

We performed a first exploration of full numerical evolutions of
a black-hole binaries with large initial separations in order to evaluate
how numerical techniques developed for close binaries
(initial separations of around $R\sim10M$) handle this regime.
We studied a prototypical binary
with an initial separation of $D=100M$. Given the large orbital period
involved ($T\approx6400M$), we restricted our evolutions to the first
few orbits. We find that the full numerical simulations
agree with the post-Newtonian predictions
for the gravitational waveform,  orbital frequencies, orbital
decay rate, and radiated energy. These are nontrivial features 
given the length scales involved in the problem and the very 
small amplitude of the radiation.

From these first studies we can draw several conclusions: \\
i) The initial pulse of spurious radiation (and gauge relaxation)
still requires a period of the order of one orbital cycle to settle
(that seems to be quite independent of the initial orbital radii).
This implies longer evolution times to obtain accurate waveform
information.  Alternatively, one could use initial data with some
information of the realistic radiation content along the lines of
Ref.~\cite{Kelly:2009js}.\\
ii) Given the long evolution times required and the long wavelengths
involved, the location of the computational boundaries should allow
for the extraction of radiation at (at least) one 
wavelength from the sources. 
We note that more efficient techniques to treat the evolution in this
{\it far} zone 
include the use of
more accurate boundary conditions~\cite{Nunez:2009wn}, multi-patch
schemes~\cite{Pollney:2009yz},
and the choice of coordinates that are better adapted to the problem
\cite{Baker:2001sf}.\\
iii) A pure 3.5PN evolution indicates that this binary will take
approximately $t\sim 8.2\times10^6M$ to inspiral to an orbital
separation of $D=5M$. During this inspiral, the binary would
complete 2064 orbits. Our full numerical simulations on 20 nodes
(3.46GHz dual Intel processors with 6 cores each) produced an average
evolution of nearly $100M$ per day. This would led to a total
of over 200 years to complete the evolution.
Dramatic  improvements in the speed of the evolution codes,
possibly using use hardware accelerators~\cite{Zink:2011pc}, or
novel numerical techniques, such as implicit-explicit
methods~\cite{Lau:2011we}, will be needed in order to make simulations
from these separations to mergers possible.\\
iv) Based on the \genb results and our results
in~\cite{Zlochower:2012fk}, we can foresee a \genc set of runs
that use the current runs to reduce eccentricity
using for instance the method \cite{Pfeiffer:2007yz},
have the computational boundaries moved to even larger radii
(which would require setting $\sigma_\infty$ to a smaller
value), use higher-order AMR prolongation and numerical dissipation,
and replace the \spd with the proper distance
of the shortest geodesic joining the two horizons.

We note that the eccentricity reduction method
of~\cite{Pfeiffer:2007yz},
when applied to $\dot s(t)$ for the $D=100M$ configuration,
gives a very small change in the initial
tangential momentum and a very large change in the initial radial
momentum.
In particular, we find $\delta p_t/p_t = 6.056\times 10^{-6}$ and
$\delta p_r/p_r = 18.6$.
The latter result is
surprising as it is not consistent with the magnitude of the PN radial
momentum. We speculate that this indicates that the oscillations in
$\dot s(t)$ are not due to purely eccentricity effects, but have a
significant gauge component, as well. If one were to modify the 
PN momentum as predicted here, and use these data as the starting
point for a 3.5PN evolution, the resulting binary would have
an initial eccentricity of $5.5\times10^{-4}$, which is roughly
the eccentricity we measure for the numerical binary.

In conclusion we have shown that we can use 
current numerical techniques
to accurately model the quasi-adiabatic
evolution of black-hole binaries
 at radii of the order of $D=100M$ and generate the 
gravitational waveforms from these
binaries, but the speed of the time integration techniques
will need to be improved by two orders of magnitude 
before such simulations can
be routine.

\acknowledgments 

The authors thank M.Campanelli and H.Nakano for discussions on this paper.
The authors gratefully acknowledge the NSF for financial support from Grants
PHY-1212426, PHY-1229173,
AST-1028087, PHY-0929114, PHY-0969855, PHY-0903782, OCI-0832606, and
DRL-1136221,  and NASA for financial support from NASA Grant No.
07-ATFP07-0158. Computational resources were provided by the Ranger
system at the Texas Advance Computing Center (XSEDE allocation
TG-PHY060027N), which is supported in part by the NSF, and by
NewHorizons and BlueSky 
at Rochester Institute of Technology, which were supported
by NSF grant No. PHY-0722703, DMS-0820923, AST-1028087, and PHY-1229173.

\bibliographystyle{apsrev}
\bibliography{../../Bibtex/references}

\end{document}